\documentclass[aps,pra,reprint,showpacs,floatfix,twocolumn,superscriptaddress]{revtex4-1}
\usepackage{amsmath,amsthm,amsfonts,amssymb,amscd}
\usepackage[utf8]{inputenc}
\usepackage[english]{babel}
\usepackage{babel}
\usepackage{graphics}
\usepackage{graphicx}
\usepackage{textcomp}
\usepackage{latexsym}
\usepackage{enumerate}
\usepackage[caption=false]{subfig}
\usepackage[colorinlistoftodos, portuguese, textwidth=140pt]{todonotes}
\usepackage{hyperref}
\usepackage{mathtools}
\usepackage{blindtext}
\usepackage{bbm}

\begin{document}
\title{Quantum-controlled cluster states}

\author{R. F. Barros}
\affiliation{Instituto de F\'isica, Universidade Federal Fluminense, CEP 24210-346, Niter\'oi-RJ, Brazil}
\email{rafael.fprb@gmail.com}
\author{G. B. Alves}
\affiliation{Instituto de F\'isica, Universidade Federal Fluminense, CEP 24210-346, Niter\'oi-RJ, Brazil}
\author{O. Pfister}
\affiliation{Department of Physics, University of Virginia, Charlottesville, Virginia 22903, USA}
\author{A. Z. Khoury}
\affiliation{Instituto de F\'isica, Universidade Federal Fluminense, CEP 24210-346, Niter\'oi-RJ, Brazil}

\begin{abstract}

Quantum optical cluster states have been increasingly explored, in the light of their importance for measurement-based quantum computing. Here we set forth a new method for generating quantum controlled cluster states: pumping an optical parametric oscillator with spatially structured light. We show that state-of-the-art techniques for producing clusters in the spectral and temporal domains are improved by a structured light pump, which manipulates the spatial mode couplings in the parametric interaction. We illustrate the method considering a second-order pump structure, and show that a simple mode rotation yields different cluster states, including, but not limited to the two-dimensional square and hexagonal lattices. We also introduce a novel technique for generating non-uniform cluster states, and propose a simple setup for its implementation. 

\end{abstract}

\maketitle

\section{Introduction}
Practical quantum optical computers have become closer to reality after recent developments with optical parametric oscillators (OPO). The OPO is a versatile device capable of generating large-scale entanglement in either time, frequency, or spatial domains. In the time domain, for example, a scheme with multiple degenerate OPOs and time-shifted Einstein-Podolsky-Rosen (EPR) pairs was shown to produce cluster states with up to $\sim 10^6$ entangled modes, with a few available at a time \cite{Yokoyama:13,Yoshikawa:16}. More recently, the same approach was used to generate a square lattice cluster state, a universal resource for one-way quantum computing \cite{Asavanant:19, Larsen:19}.

An alternative approach uses a single non-degenerate OPO and a spectrally shaped pump beam to tailor the interaction between shifted frequency combs  -- also known as  a quantum optical frequency combs (QOFC). Such device was shown to produce different outcomes, be it in the form of parallel copies of low-dimensional states \cite{Pysher:11} or large-scale cluster states \cite{Chen:14,Menicucci:08}. More recently, a novel scheme was proposed to produce square lattice cluster states via phase-modulation of the QOFC \cite{Zhu:20}, which is promising for integrated quantum photonics. Nonetheless, a singular technique of synchronously pumping the OPO with a pulsed laser was also proven to produce cluster states \cite{Roslund2014,Cai2017}, with modes encoded in the temporal shape of the downconverted fields.

On a lower scale, the spatial degree of freedom was also shown to improve the capabilities of currently available systems by increasing the dimensionality of the parameter space. For example, one could mention the production of multipartite entanglement between first-order spatial modes in a type-II OPO \cite{Coutinho:09,Liu:14,Liu:16,Cai:18}. In a series of recent reports, several cluster states have also been produced with orbital angular momentum (OAM) modes in a four-wave mixing setup \cite{Pooser:14,Pan:19,Li:20,Zhang:20,Liu:20,Wang:20}, with applications to quantum teleportation and quantum networks. One should also mention the proposals for production of both spatial-only \cite{Zhang:17} and spatio-spectral \cite{Yang:20} graph states with OAM modes in the OPO.

In this paper, we investigate the role of the pump spatial structure in the two main techniques for generating cluster states with OPOs, namely the frequency and time shifting of quantum combs. We show that the nonlinear interaction between higher order spatial modes can be easily manipulated with the pump structure, allowing for the versatile control of the states produced by the aforementioned techniques. We illustrate the method considering a second-order pump mode, showing that it leads to quantum controlled one-dimensional (1D) clusters in the spectral domain. Furthermore, we show that time shifting the 1D spectral clusters leads to different two-dimensional (2D) clusters in the hybrid spectro-temporal domain, including, but not limited to the square and hexagonal topologies. Lastly, we propose a novel technique for generating non-uniform 2D clusters, exploring the rapidly switching pump structure obtained by polarization modulation of classically non-separable vector modes.

The paper is organized as follows. In section \ref{formalism} we introduce the formalism we use to derive the quantum states produced by the OPO. In section \ref{quadripartite} we analyze the tuning of the quadripartite interaction enabled by a varying second-order spatial mode in the pump field. This is the building block we use to implement quantum control over spectral clusters in section \ref{1d} and spectro-temporal clusters in sections \ref{2d} and \ref{non-uniform}. Lastly, in section \ref{experiment} we comment on the possible experimental challenges of the proposed setups. 

\section{Formalism}\label{formalism}

The Hamiltonian of the optical parametric interaction, under the undepleted pump approximation, is the following \cite{Patera:12}
\begin{equation}
\hat{H}=i\hbar\kappa\sum_{mn} G_{mn}\left(\hat{a}_m^\dagger\hat{a}_n^\dagger -\hat{a}_m\hat{a}_n\right)\,, 
\label{eq-hamiltonian}
\end{equation}
where $\hat{a}_{m,n}^\dagger$ ($\hat{a}_{m,n}$) are the creation (annihilation) operators of the downconverted modes and $\kappa$ is a coupling strength, which is proportional to the second-order susceptibility. The matrix $\mathbf{G}$ is the adjacency matrix of the Hamiltonian graph \cite{Menicucci:07,Menicucci:11}, which in the case of perfect phase-matching can be written as
\begin{equation}
G_{mn}=\sum_l\alpha_l \Lambda_{lmn}\,.
\label{adjacenc-matrix}
\end{equation}
It depends on the modal content of the pump via the amplitudes $\alpha_l$, and on the coupling between pump ($p$), signal ($s$) and idler ($i$) transverse modes via the overlap integral $\Lambda_{lmn}=\int d^2r\, u_l^{(p)}(\mathbf{r})u_m^{(s)*}(\mathbf{r})u_n^{(i)*}(\mathbf{r})$. The overlap integral contains the selection rules of the transverse mode coupling, as extensively discussed in Refs. \cite{Alves2018,Schwob1998}.


Our goal is to investigate the production and manipulation of multipartite entanglement using the pump spatial structure as resource. For this purpose, we consider the case of a second order pump mode, whose electric field can be generally written in the Hermite-Gaussian basis ($\textrm{HG}_{mn}$) \cite{Yariv} as $E_p=\alpha_{20}\textrm{HG}_{20} + \alpha_{11}\textrm{HG}_{11}+\alpha_{02}\textrm{HG}_{02}$, coupling to first order modes in the signal and idler. According to the selection rules detailed in Ref. \cite{Alves2018, Schwob1998}, three concurrent nonlinear processes are allowed in this case, namely HG$_{02}\!\!\rightarrow$HG$_{10}$+HG$_{10}$, HG$_{02}\!\!\rightarrow$HG$_{01}$+HG$_{01}$ and HG$_{11}\!\!\rightarrow$HG$_{10}$+HG$_{01}$. Such concurrent interactions are known to yield quadripartite CV entangled states, as experimentally verified in \cite{Liu:16}.

In the next section, we explore the control over the aforementioned multimode entangled states by tuning the interaction Hamiltonian \eqref{eq-hamiltonian} with a varying pump spatial structure. 

\section{Quadriparite cluster states from a spatially structured pump}\label{quadripartite}


Let us consider a convenient example of pump structure, defined by the amplitudes $\alpha_{11}=\sin2\theta$ and $\alpha_{20}=-\alpha_{02}=\cos2\theta/\sqrt{2}$. This corresponds to a HG$_{11}$ mode rotated clockwise by $\theta-\pi/4$, whose electric field can be written in the Hermite-Gaussian basis \cite{Yariv} as $E_p=\cos2\theta\,(\textrm{HG}_{02} -\textrm{HG}_{20})/\sqrt{2} + \sin2\theta\,\textrm{HG}_{11}$. Naming the first-order downconverted modes as $\{\hat{a}_1,\hat{a}_2,\hat{a}_3,\hat{a}_4\}=\{\hat{a}^{(s)}_{10},\hat{a}^{(s)}_{01},\hat{a}^{(i)}_{10},\hat{a}^{(i)}_{01}\}$, with $s$ and $i$ standing for signal and idler, respectively, the adjacency matrix \eqref{adjacenc-matrix} becomes
\begin{eqnarray}
\mathbf{G}(\theta)&=&
\begin{pmatrix}
\mathbf{0}&\boldsymbol{\alpha}\\
\boldsymbol{\alpha}&\mathbf{0}
\end{pmatrix}\,,
\label{G-matrix}\\
\boldsymbol{\alpha}&=&
\frac{1}{\sqrt{2}}
\begin{pmatrix}
\cos 2\theta &\sin 2\theta\\
\sin 2\theta &-\cos 2\theta
\end{pmatrix}\,,
\label{eq-alpha}
\end{eqnarray}
up to a multiplicative constant that can be absorbed into the coupling strength $\kappa$. The action of such multimode squeezing interaction in phase space is fully described by a symplectic matrix (see Appendix \ref{appendix-symplectic}), which in the present case is given by $\mathbf{S}=e^{\gamma\mathbf{M}}$, with $\mathbf{M}=\textrm{diag}[\mathbf{G},-\mathbf{G}]$ and $\gamma=2\kappa t$ for an interaction time $t$. From $\mathbf{S}$ we readily obtain the covariance matrix of the transformed quantum state, to which we apply the Simon criterion \cite{Simon:00} to identify entanglement. 

\begin{figure}
\includegraphics[scale=0.5]{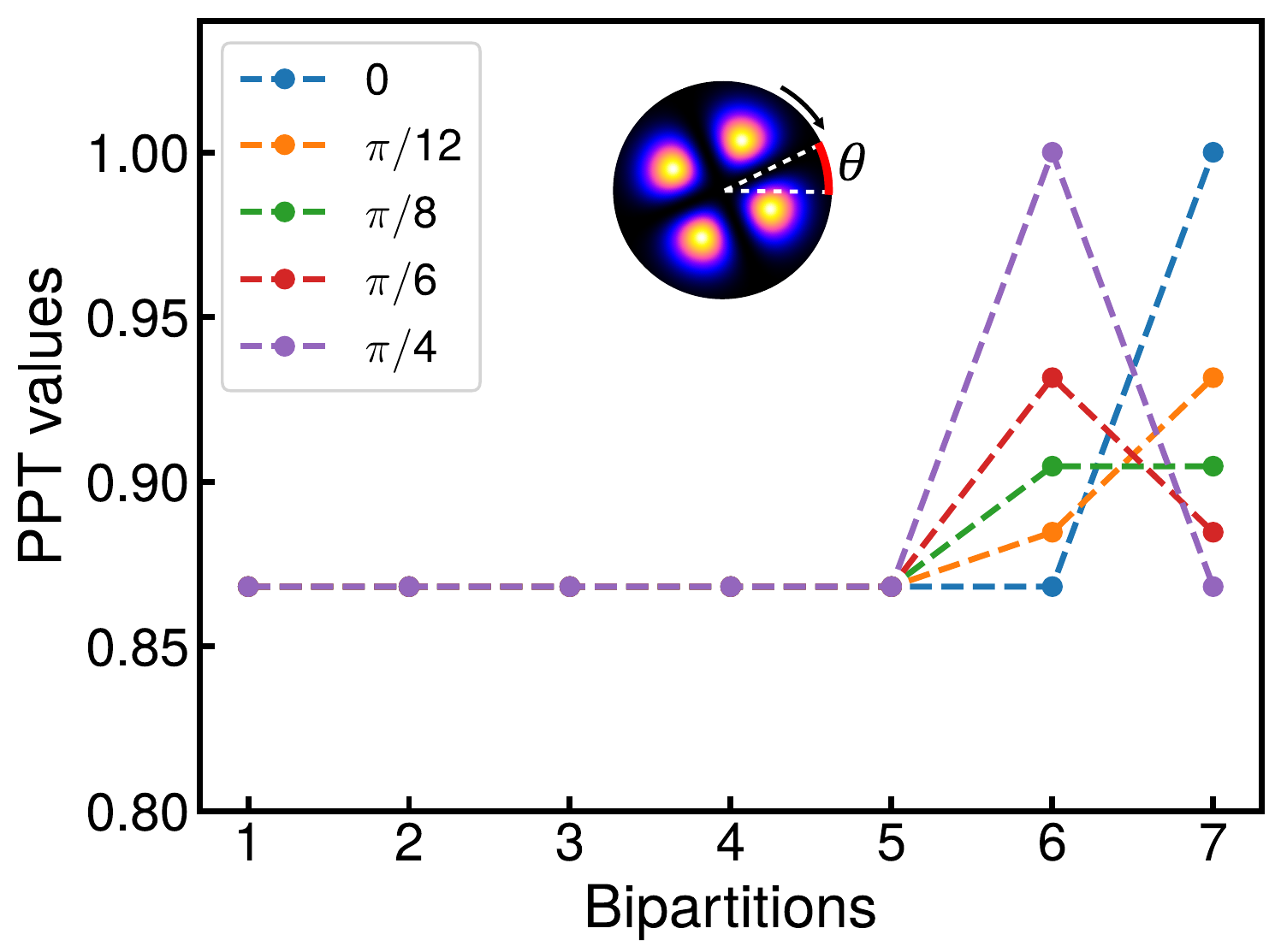}
\caption{PPT values obtained for a HG$_{11}$ pump mode rotated clockwise by $\pi/4-\theta$. Each curve corresponds to a different value of $\theta$, and we considered $\gamma=0.1$.}
\label{Fig-symplectic}
\end{figure}

We show in Fig.\ref{Fig-symplectic}a the lowest symplectic eigenvalues (PPT values) of the partially transposed covariance matrices obtained for the adjacency matrix \eqref{G-matrix}. The PPT values are shown for the seven possible bipartitions, namely ($1$) $1|234$, ($2$) $2|134$, ($3$) $3|124$, ($4$) $4|123$, ($5$) $12|34$, ($6$) $13|24$ and ($7$) $14|23$, for different values of the rotation angle $\theta$. Notice that for all angles between $0<\theta<\pi/4$, all the PPT values are lower than one, indicating genuine quadripartite entanglement . The quadripartite entanglement is broken only for $\theta=0$ or $\theta=\pi/4$, in which cases at least one of the three HG components is absent.

\begin{figure*}
\includegraphics[scale=0.4]{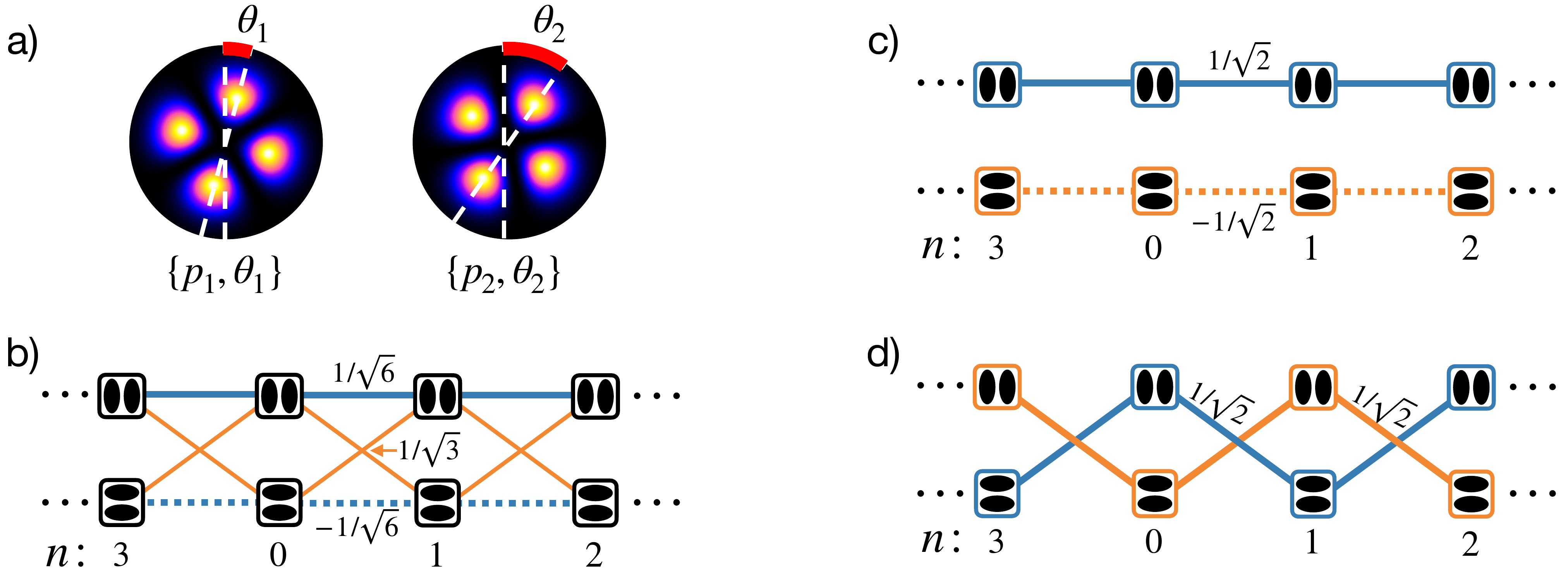}
\caption{Cluster states produced by dual-frequency spatially structured pump. In (a) we show the considered pump modes, which are rotated petal modes. The produced cluster states are shown for $p_1=1$ and $p_2=3$, and the following combinations of spatial modes: (b) $\theta_1=\theta_2=\pi/8$, (c) $\theta_1=\theta_2=0$ and (d) $\theta_1=\theta_2=\pi/4$. While in (b) the cluster has a dual-rail structure, in (c) and (d) it is dismantled in two independent single-rail clusters. The numbers in (b), (c) and (d) are the weights of the cluster edges. }
\label{Fig-clusters}
\end{figure*}

From the symplectic formalism we also get the squeezed supermodes -- combinations of field quadratures that become squeezed in the interaction --, given by the negative eigenvectors of the matrix $\mathbf{M}$. They are
\begin{equation}
\begin{aligned}
\left[-\hat{Q}_1\cos 2\theta -\hat{Q}_2\sin2\theta +\hat{Q}_3\right]\propto e^{-\gamma/\sqrt{2}} \,,\\
\left[\hat{Q}_1\sin 2\theta -\hat{Q}_2\cos 2\theta -\hat{Q}_4\right]\propto e^{-\gamma/\sqrt{2}} \,,\\
\left[\hat{P}_1\cos2\theta +\hat{P}_2\sin2\theta +\hat{P}_3\right]\propto e^{-\gamma/\sqrt{2}} \,,\\
\left[\hat{P}_1\sin2\theta -\hat{P}_2\cos2\theta +\hat{P}_4\right]\propto e^{-\gamma/\sqrt{2}} \,.
\end{aligned}
\label{squeezed-quadripartite}
\end{equation}
These combinations of field quadratures, after appropriate phase shifts, are the approximate nullifiers of a quadripartite cluster state whose adjacency matrix is precisely $\mathbf{G}$, due to the fact that $\mathbf{G}^2=\mathbbm{1}$ \cite{Menicucci:07}.

In the following sections, we will use the aforementioned quadripartite system as a building block to perform quantum control over large-scale cluster states produced by state-of-the-art techniques, namely the quantum optical frequency comb (QOFC) and the time-shifting with delay lines.

\section{Spatio-spectral cluster states}\label{1d}

A QOFC \cite{Menicucci:08,Pysher:11,Chen:14} is usually obtained from a type-I (or type-0) OPO operating below threshold and close to degeneracy, which features several longitudinal modes simultaneously close to resonance \cite{Eckardt:91,Barros:20}. The downconverted frequencies are $\omega_n=\omega_0+n\delta$, where $\omega_0$ is an arbitrary offset and $\delta$ is the cavity free spectral range. Due to energy conservation, the pair $\{n_1,n_2\}$  only couples to a pump frequency given by $\omega_p=2\omega_0+p\delta$, where $p=n_1+n_2$ is the pump index. Therefore, each spectral component $p$ of the pump produces multiple EPR pairs dispersed in the frequency domain, fundamentally limited by the phase-matching bandwidth of the nonlinear crystal \cite{Wang:14}. When two pump frequencies are used, the EPR pairs concatenate and form a large-scale 1D cluster state \cite{Chen:14}.

\begin{figure*}
\includegraphics[scale=0.6]{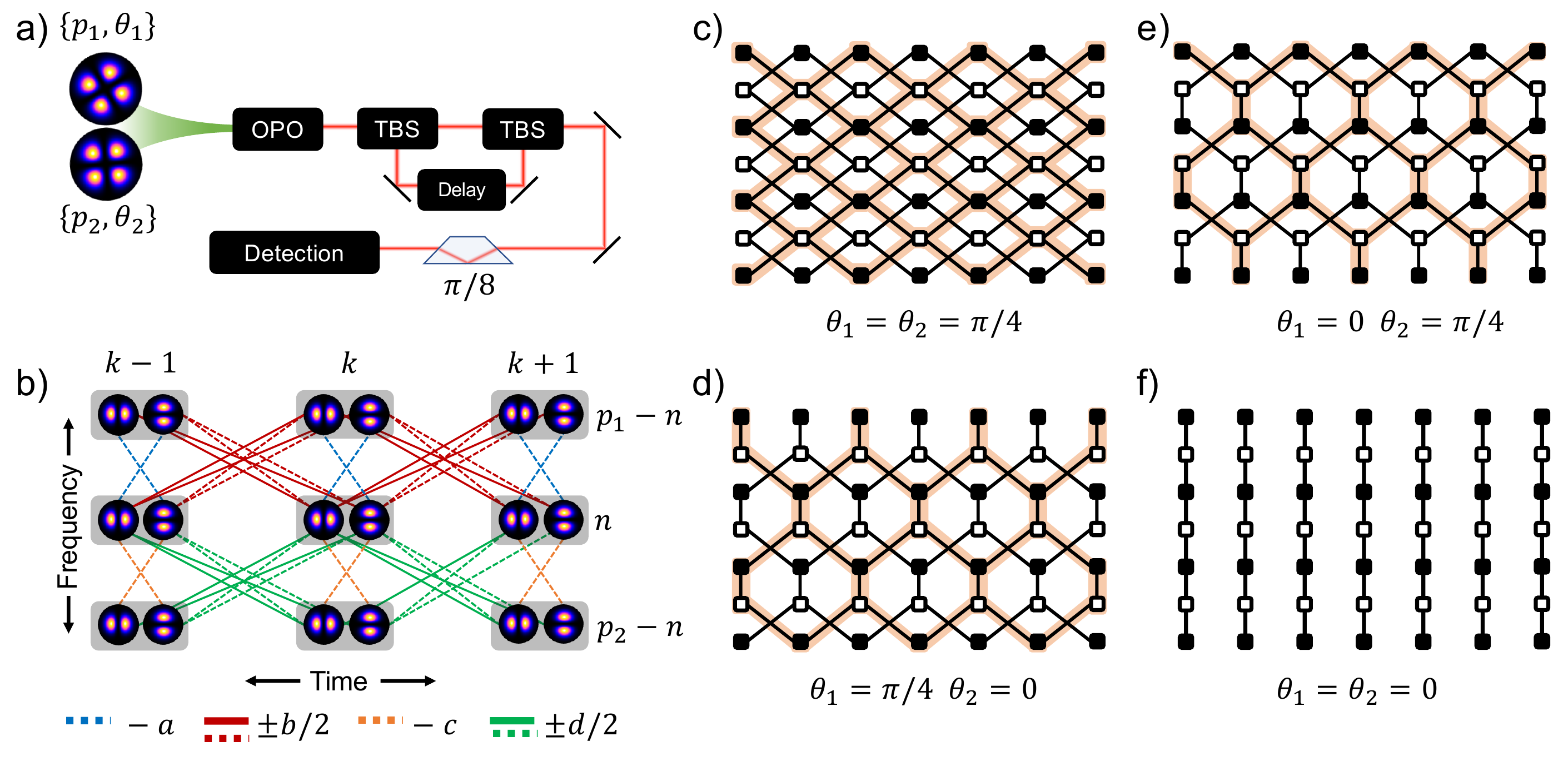}
\caption{(a) Experimental apparatus for the production of a bilayer 2D lattice with structured light. The detailed cluster graph is shown in (b). In panels (c)-(f), we show the resulting clusters for different pump structures, defined by the the rotation angles $\theta_1$ and $\theta_2$ indicated in each panel. The filled and unfilled squares indicate frequency indices $n$ with different parities, and the shaded regions in (c)-(e) highlight the 2D structures achieved in each case.}
\label{Fig-2D}
\end{figure*}

Here we consider a dual-frequency pump with a rotated HG$_{11}$ mode in each frequency component, which is labeled by the indices $\{p_1,p_2\}$ for the frequencies and $\{\theta_1,\theta_2\}$ for the rotation angles, as shown in Fig.\ref{Fig-clusters}a. While each pump spectral component produces several quadripartite entangled states, both combined yield a cluster state known as a dual-rail quantum wire \cite{Chen:14,Yokoyama:13}, shown in Fig.\ref{Fig-clusters}b for $\theta_1=\theta_2=\pi/8$ (see Appendix \ref{appendix-dual-frequency} for detailed calculations).

Notice that each cluster node has a given frequency and spatial mode, while the edges are controlled by the rotation angles of the pump modes. This is exemplified in Figs.\ref{Fig-clusters}c and \ref{Fig-clusters}d, which show the extreme cases in which at least one of the pump HG components is absent. In these cases, the dual-rail structure is dismantled in two independent single-rails, connecting identical (orthogonal) spatial modes for $\theta_1=\theta_2=0$ ($\theta_1=\theta_2=\pi/4$). These particular examples already show that pumping an OPO with structured light provides an efficient way to implement quantum control over spectral cluster states.

\section{Extension to universal cluster states}\label{2d}

The dual-rail cluster state, although highly scalable, is one-dimensional and thus not suitable for universal for MBQC \cite{Gu:09}. However, it is possible to create a two-dimensional (2D) cluster by mixing several dual-rail clusters emitted sequentially by the OPO \cite{Alexander:16}. When sufficiently squeezed \cite{Menicucci:14} and used alongside non-Gaussian resources such as Gottesman-Kitaev-Preskill (GKP) states and photon-number resolving measurements \cite{Lloyd:99,Braunstein:05}, a 2D cluster state is suitable for fault-tolerant and universal MBQC. In this section we discuss how the strategy of Ref.\cite{Alexander:16} can be applied to spatially structured fields and how the quantum control over the primary dual-rail clusters manifests in the resulting universal resource.

We illustrate in Fig.\ref{Fig-2D}a the proposed experimental setup, which is the spatial-mode equivalent of the scheme reported in Ref.\cite{Alexander:16}. First, the OPO is pumped by the continuous-wave spatio-spectrally structured field detailed in the previous section and produces a stream of the dual-rail clusters shown in Fig.\ref{Fig-clusters}. Although the outgoing fields are also continuous-wave, they have a coherence time on the order of the OPO lifetime, measured as the inverse of the squeezing bandwidth. This allows one to divide the downconverted fields in independent ``time bins'' or temporal modes $k\in\mathbb{Z}$, which are emitted sequentially by the OPO \cite{Asavanant:19,Yokoyama:13,Yoshikawa:16}. 

The spatio-spectral clusters in each temporal index $k$ carry first-order HG spatial modes, which are separated in a transverse mode beam splitter (TBS) \cite{Sasada:03,Passos:20}. The vertically oriented HG$_{01}$ modes are then time-delayed by one temporal index and reunited with the HG$_{10}$ modes. The last element before the detection is a Dove prism, which performs a $\pi/4$ clockwise rotation on the spatial modes.

The graph of the resulting 2D cluster state is shown in Fig.\ref{Fig-2D}b, and detailed calculations of the corresponding nullifiers are provided in Appendix \ref{appendix-2D}. The cluster is a macronode-based bilayer lattice \cite{Alexander:16, Larsen:19}, with each macronode containing a pair of orthogonal HG modes with the same spectral and temporal indices. The reconfigurability of the 2D cluster is evidenced by the dependence of the cluster edges on the pump structure, contained in the following parameters 
\begin{equation}
\begin{aligned}
a=\frac{\cos2\theta_1}{2r}\,,\qquad b=\frac{\sin2\theta_1}{\sqrt{2}r}\,,\\
c=\frac{\cos2\theta_2}{2r}\,,\qquad d=\frac{\sin2\theta_2}{\sqrt{2}r}\,,
\end{aligned}
\label{rail-edges}
\end{equation}
where $r$, given by
\begin{equation}
r=\frac{\sqrt{6-\cos4\theta_1-\cos4\theta_2}}{2\sqrt{2}}\,.
\label{rail-squeezing}
\end{equation} 
We show in Fig.\ref{Fig-2D}c-f the graphs of the resulting clusters for some combinations of pump angles. 

The quantum control feature displayed by the dual-rails of Fig.\ref{Fig-clusters} is clearly inherited by the 2D structures obtained with the temporal encoding. For $\theta_1=\theta_2=\pi/4$, as highlighted in Fig.\ref{Fig-2D}c, the OPO produces two independent \textit{bilayer square lattices} \cite{Alexander:16}, connecting macronodes with neighbouring temporal indices and frequency indices of different parities. On the other hand, for $\theta_1=0$ and $\theta_2=\pi/4$ ($\theta_1=\pi/4$ and $\theta_2=0$), we show in Fig.\ref{Fig-2D}d (\ref{Fig-2D}e) that the produced clusters are \textit{bilayer hexagonal lattices}, which are also universal for MBQC \cite{Nest:06}. Finally, for $\theta_1=\theta_2=0$, the connections between different temporal modes vanish and the cluster loses its 2D character. These results show that, with the proposed strategy, cluster transformations which would require cumbersome sets of local measurements can be performed by a single change in the pump spatial structure.

\section{Time-varying pump structure}\label{non-uniform}

\begin{figure}
\includegraphics[scale=0.6]{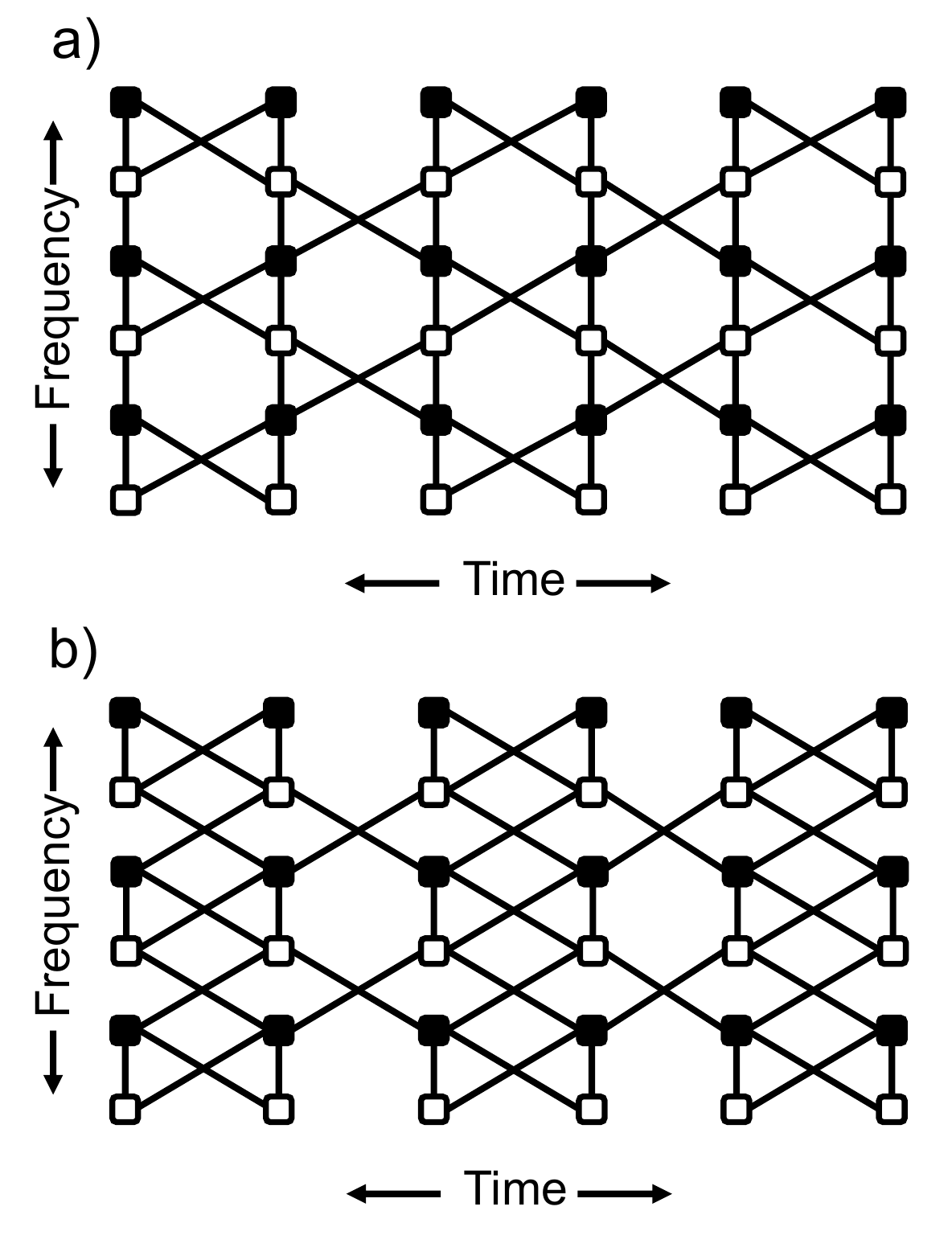}
\caption{Graphs for cluster states produced by a pump structure $\{\theta_1,\theta_2\}$ varying between (a) $\{0,\pi/4\}$ and $\{\pi/4,0\}$ (b) $\{0,\pi/4\}$ and $\{\pi/4,\pi/4\}$.}
\label{Fig-onthefly}
\end{figure}

The 2D clusters in the aforementioned discussion are built by time-shifting a sequence of identical spatio-spectral 1D clusters. An additional layer of quantum control is achieved if the pump structure is such that the OPO produces a sequence of different 1D clusters instead. To illustrate this phenomenon, we show in Fig.\ref{Fig-onthefly} the resulting clusters for a pump field switching back and forth between two configurations on a time scale shorter than the down-converted time-bins, according to the nullifiers derived in Appendix \ref{appendix-2D}. Note that two completely different structures are produced in this case. while in Fig.\ref{Fig-onthefly}a we have a Kagome-like pattern, the graph of Fig.\ref{Fig-onthefly}b displays a non-uniform pattern. Naturally, this additional layer of quantum control can be further explored by considering different sequences of three or more pump spatial structures, using the 2D building blocks shown in Fig.\ref{Fig-2D}.  

\begin{figure}
\includegraphics[scale=0.8]{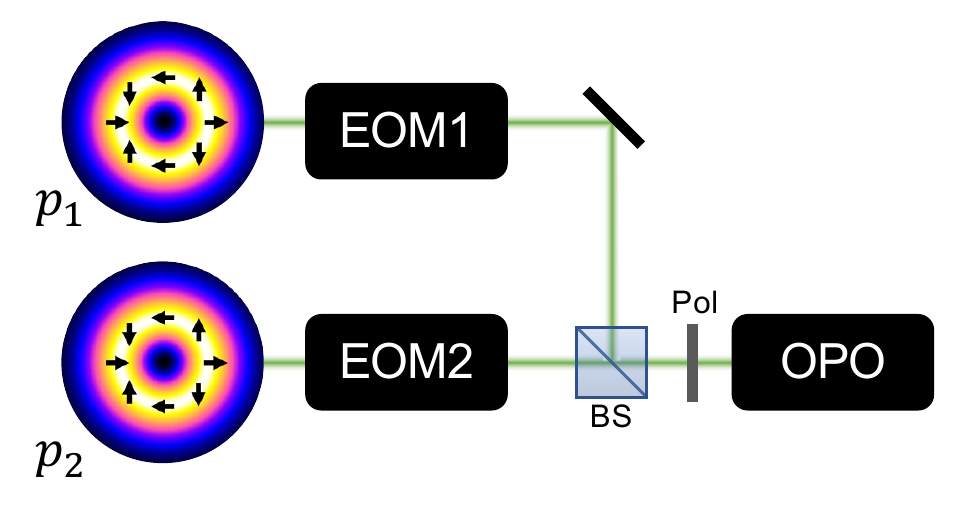}
\caption{Simplified experimental arrangement to produce separate time-varying spatial structures for the two pump frequency components. The pump frequency modes $p_1$ and $p_2$ carry second-order vector vortices in their transverse structures, and the spatial mode switching is achieved by passing each of them through a polarization EOM, a non-polarizing beam-splitter (BS) and a polarizer (Pol). }
\label{Fig-vortex-setup}
\end{figure}

We also propose a simple experimental arrangement for the fast switching between different pump structures, displayed in Fig.\ref{Fig-vortex-setup}. First, each pump spectral component is prepared in a second-order vector vortex mode, whose electric field is given by $\mathbf{E}=(\textrm{HG}_{20} - \textrm{HG}_{02})\mathbf{e_h}/\sqrt{2}+\textrm{HG}_{11}\mathbf{e_v} $, where $\mathbf{e_h}$ and $\mathbf{e_v}$ denote horizontal and vertical polarizations, respectively. Then the beams pass through polarization electro-optical modulators (EOM), that switch the polarizations in MHz frequencies, and through a polarizer. Due to the non-separability between spatial structure and polarization of such modes, the outcome of the polarizer will be a superposition of second-order petal modes with different frequencies, switching between angles controlled by the EOMs. 

As the EOM frequencies easily surpass the squeezing bandwidth of typical OPOs, we can assume that the pump modes are constant for the duration of a time bin and change sharply at the end. Such square wave changes in the pump manifest equivalently in the interaction Hamiltonian, which then becomes a piece-wise constant function of time, different for each temporal index.  This results in a sequence of dual-rail clusters that can be directly controlled by the waveforms applied to the EOMs.

\section{Experimental feasibility}\label{experiment}
While the production and manipulation of large-scale cluster states in the spectral and temporal domains are already well established techniques, adding the spatial degree of freedom may bring additional challenges. The most prominent bottleneck would be the astigmatism introduced by the nonlinear crystal \cite{Martinelli:04}, which would separate the frequency combs for the modes HG$_{10}$ and HG$_{01}$. While a solution for this problem has been reported  \cite{Liu:14}, its application to large-scale clusters remains to be tested.

The characterization of the produced state can also be challenging, as standard homodyne measurements require independent local oscillators in all the different spatial modes and frequencies. This may be critical specially if higher-order modes are used. One possible path to overcome this limitation is the parametric homodyne technique \cite{Shaked:18}, which eliminates the need for local oscillators and makes the squeezing measurements robust to noise.  Associated with modern mode-sorting strategies \cite{Morizur:10,Fontaine:19,Fickler:20}, this technique is promising for MBQC with spatially structured clusters.

\section{Conclusion}
We have presented a novel method to control continuous-variable cluster states produced by OPOs, exploring the spatial structure of the pump field. Considering second order pump modes, we have shown that reconfigurable 1D cluster states are produced in the spectral domain, extending to universal 2D clusters after time shifting with a delay line. Nonetheless, the 2D clusters inherit the reconfigurability of the 1D resource, and can assume different topologies as the pump field is rotated. We have also proposed a novel scheme for generating non-uniform 2D clusters using a rapidly switching pump structure, and presented a simple experimental scheme for its implementation using non-separable vector beams and polarization modulators. Our results open a new path for the production and control of large-scale cluster states, settling the importance of the spatial degree of freedom for measurement-based quantum computing.

\appendix

\section{Symplectic formalism}\label{appendix-symplectic}

In this appendix we introduce the symplectic formalism that we use in our entanglement analysis. First, we define the quadrature vector $\mathbf{\hat{x}}=(\hat{Q}_1\,\hat{Q}_2\,...\,\hat{Q}_n\, \hat{P}_1\,\hat{P}_2\,...\,\hat{P}_n)$, for which the symmetrized covariance matrix is given by \cite{Menicucci:11}
\begin{equation}
\mathbf{V}=\textrm{cov}\,\mathbf{\hat{x}}=\frac{1}{2}\langle \{\mathbf{\hat{x}}^\dagger,\mathbf{\hat{x}}^T\} \rangle\,.
\end{equation}
In the above equation, we used the following definition for the anti-commutator 
\begin{equation}
\{\mathbf{\hat{r}},\mathbf{\hat{s}}^T\}:=\mathbf{\hat{r}} \mathbf{\hat{s}}^T + (\mathbf{\hat{s}} \mathbf{\hat{r}}^T)^T\,,
\label{anti-commutator}
\end{equation}
where $\mathbf{\hat{r}}$ and $\mathbf{\hat{s}}$ are operator-valued vectors, and the Hermitian conjugation only applies to the operators within the vectors \cite{Menicucci:11}. Notice that, with our definitions for the field quadratures ($\hat{Q}=\hat{a}+\hat{a}^\dagger$ and $\hat{P}=-i(\hat{a}-\hat{a}^\dagger)$), the covariance matrix of the vacuum is simply $\mathbf{V}=\mathbbm{1}$, where $\mathbbm{1}$ is the identity matrix.

Now we turn to the Heisenberg time evolution of the quadrature vector $\mathbf{\hat{x}}$ under a Gaussian Hamiltonian -- a Hamiltonian at most quadratic in the field operators. It can be shown that \cite{Simon:88}
\begin{equation}
\mathbf{\hat{x}}(t)=\mathbf{S}\mathbf{\hat{x}}(0)\,,
\end{equation}
where $\mathbf{S}=\mathbf{S}(t)$ is a symplectic matrix of complex numbers and $t$ is the time. In the special case where the initial state is vacuum, the covariance matrix after the Gaussian interaction is simply given by \cite{Menicucci:11}
\begin{equation}
\mathbf{V}=\mathbf{S}\mathbf{S}^T\,,
\label{covariance}
\end{equation}
which is the expression we use in the following analysis. 

There are different criteria to identify the entanglement of a Gaussian state, but here we focus on the one proposed by Simon \cite{Simon:00}. It consists of a generalization to continuous variables of the Peres-Horodecki (PPT) criterion \cite{Peres:96}, which relies on the positivity of the density matrix after partial transposition. For CV systems, the partial transposition translates into a ``local time reversal'' ($\hat{Q}_j\rightarrow\hat{Q}_j$ and $\hat{P}_j\rightarrow -\hat{P}_j$) of the transposed subsystem \cite{Simon:00}, and the condition for a physical covariance matrix is
\begin{equation}
\mathbf{V}+i\boldsymbol{\Omega}\geq 0\,,
\label{Simon-criterion}
\end{equation}
where 
\begin{equation}
\boldsymbol{\Omega}=-i[\mathbf{\hat{x}},\mathbf{\hat{x}}^T]=
\begin{pmatrix}
\mathbf{0}&\mathbf{I}\\
-\mathbf{I}&\mathbf{0}
\end{pmatrix}\,,
\end{equation}
with the commutator of operator-valued vectors defined in analogy to \eqref{anti-commutator}. If the partially transposed covariance matrix violates \eqref{Simon-criterion}, the chosen bipartition is entangled.


The violation of \eqref{Simon-criterion} can be conveniently tested by inspection of the \textit{symplectic eigenvalues}, defined as the ordinary eigenvalues of the matrix $-i\boldsymbol{\Omega}\mathbf{V}$. With our definitions, if the lowest symplectic eigenvalue of the partially transposed covariance matrix (PPT value) is lower than unit, \eqref{Simon-criterion} is violated and the selected bipartition is entangled. 

\section{Cluster nullifiers for a dual-frequency pump}
\label{appendix-dual-frequency}

Here we consider in more detail the case of a dual-frequency pump, labeled by the indices $\{p_1,\theta_1\}$ and $\{p_2,\theta_2\}$, with $p_j$ indicating the frequency and $\theta_j$ the spatial mode in that frequency. The two pumps are assumed to have the same intensity and phase, for simplicity.  The corresponding interaction Hamiltonian can be written as
\begin{equation}
\hat{\mathcal{H}}=i\frac{\kappa}{2}\hbar\sum_n \left(V^T_n\bar{\mathbf{G}}V_n-H.C\right)\,,
\label{dual-freq-Hamiltonian}
\end{equation}
where the summation is over the frequency indices $n$ of the quantum optical frequency comb (QOFC). Also, the matrices $\bar{\mathbf{G}}$ and $V_n$ are given by
\begin{equation}
\bar{\mathbf{G}}\!=\!
\begin{pmatrix}
\mathbf{G}(\theta_1)&\mathbf{0}\\
\mathbf{0}&\mathbf{G}(\theta_2)\\ 
\end{pmatrix}
\,,
V_n=
\begin{pmatrix*}[l]
\hat{b}_{p_1-n}^\dagger \\ 
\hat{c}_{p_1-n}^\dagger\\
\hat{b}_{n}^\dagger\\
\hat{c}_{n}^\dagger\\
\hat{b}_{p_2-n}^\dagger\\
\hat{c}_{p_2-n}^\dagger
\end{pmatrix*}.\!
\label{adjacency-rail}
\end{equation}
where we have renamed the operators $\hat{a}_{10}^j$ and $\hat{a}_{01}^j$ as $\hat{b}_j$ and $\hat{c}_j$, respectively, with $j$ being the frequency index. 


\begin{figure*}
\includegraphics[scale=0.83]{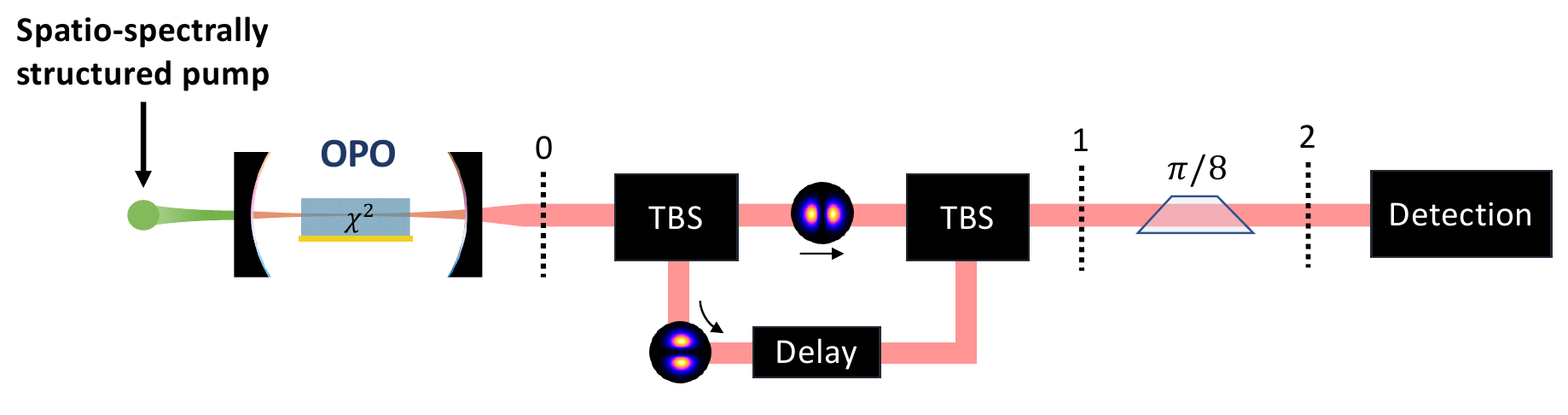}
\caption{Proposed experimental scheme for the production of 2D cluster states with structured light.}
\label{Fig-2D-setup}
\end{figure*}

From \eqref{adjacency-rail}, we obtain the corresponding symplectic matrix and extract the squeezed supermodes. In particular, we obtain that the following quantities are squeezed
\begin{equation}
\begin{aligned}
X_h\!=\hat{Q}_h^n-\{a\hat{Q}_h^{p_1-n}+b\hat{Q}_v^{p_1-n}+c\hat{Q}_h^{p_2-n}+d\hat{Q}_v^{p_2-n}\}\,,\\
X_v\!=\hat{Q}_v^n-\{b\hat{Q}_h^{p_1-n}-a\hat{Q}_v^{p_1-n}+d\hat{Q}_h^{p_2-n}-c\hat{Q}_v^{p_2-n}\}\,,
\end{aligned}
\label{dual-rail-squeezing}
\end{equation}
where the subscripts $h$ and $v$ indicate the spatial modes $HG_{10}$ and $HG_{01}$, respectively, and the parameters $a$, $b$, $c$ and $d$ are defined in Eq.\eqref{rail-edges}. These squeezed quantities are the approximate nullifiers of a one-dimensional cluster state known as a dual-rail quantum wire \cite{Chen:14}, whose graph is shown in Fig.\ref{Fig-clusters}.

\section{Time-staggering of 1D clusters states}
\label{appendix-2D}

\subsection{Fixed pump structure}

\begin{figure}[b]
\includegraphics[scale=0.7]{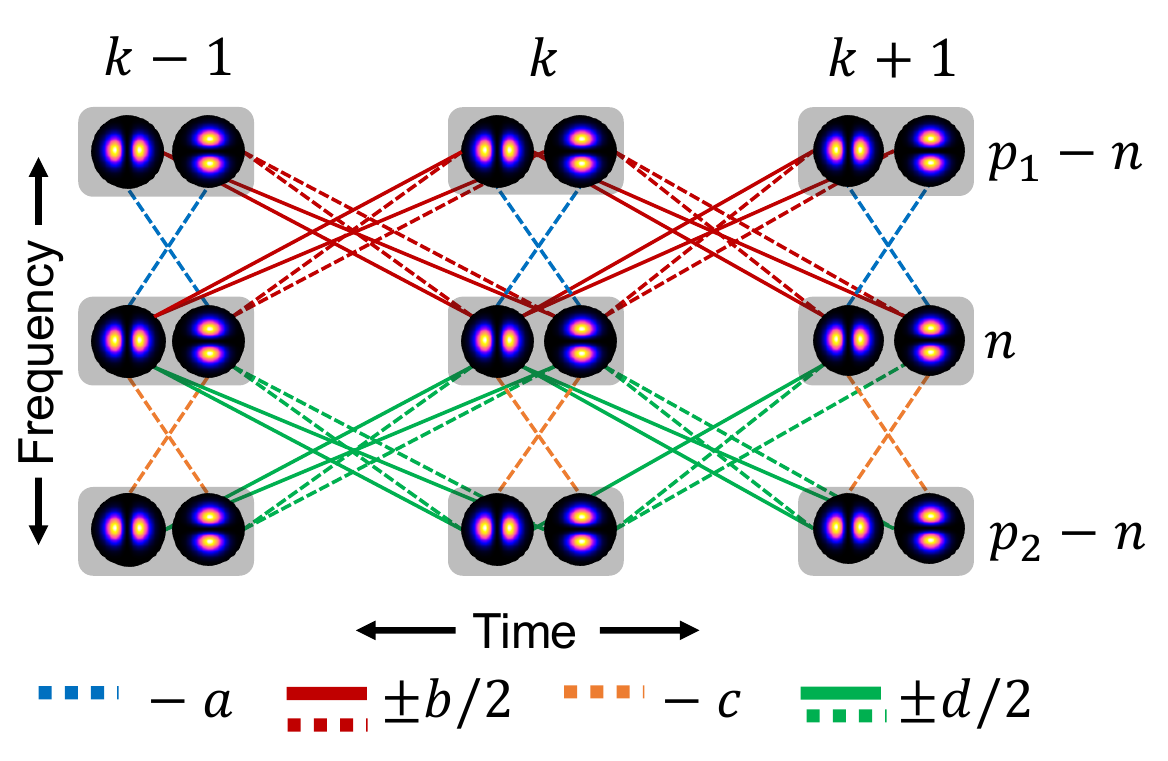}
\caption{Graph of the 2D cluster state defined by the nullifiers \eqref{nullifier_+} and \eqref{nullifier_-}.}
\label{Fig-2D-cluster}
\end{figure}

In this appendix we detail the generation of a two-dimensional (2D) cluster state from the dual-rail clusters defined in \eqref{dual-rail-squeezing}, following the strategy described in Ref.\cite{Alexander:16}. The proposed experimental setup is shown in Fig.\ref{Fig-2D-setup}. First, the OPO generates a sequence of dual-rail quantum wires arriving at point $0$, each one with a different temporal index $k$. Then the HG$_{10}$ transverse modes are delayed by one temporal index with respect to the HG$_{01}$ modes, before reaching the point $1$. The last transformation is a spatial mode rotation with a dove prism, which mixes the spatial modes arriving at $2$ at the same time. The corresponding transformations of the field quadratures from $0$ to $2$ are the following
\begin{equation}
\begin{aligned}
\hat{Q}_{h,k}^{(0)}&=\frac{1}{\sqrt{2}}\left(\hat{Q}_{h,k}^{(2)}-\hat{Q}_{v,k}^{(2)}\right)\,,\\
\hat{Q}_{v,k}^{(0)}&=\frac{1}{\sqrt{2}}\left(\hat{Q}_{h,k+1}^{(2)}+\hat{Q}_{v,k+1}^{(2)}\right)\,,
\end{aligned}
\label{transformation}
\end{equation}
where the upper indices $0$ and $2$ indicate the different stages in the setup. Note that the frequency index was omitted in the above equations, but these transformations apply independently to all the frequencies of the QOFC.

The next step is to map the linear optical transformations \eqref{transformation} on the 1D dual-rail clusters. From \eqref{dual-rail-squeezing}, we obtain
\begin{widetext}
\begin{equation}
\begin{aligned}
X_{h,k}&=\hat{Q}^{n}_{h,k}-\hat{Q}^{n}_{v,k}-\left[a(\hat{Q}^{p_1-n}_{h,k}-\hat{Q}^{p_1-n}_{v,k})+b(\hat{Q}^{p_1-n}_{h,k+1}+\hat{Q}^{p_1-n}_{v,k+1})+c(\hat{Q}^{p_2-n}_{h,k}-\hat{Q}^{p_2-n}_{v,k})+d(\hat{Q}^{p_2-n}_{h,k+1}+\hat{Q}^{p_2-n}_{v,k+1})\right]\,,\\
X_{v,k}&=\hat{Q}^{n}_{h,k+1}+\hat{Q}^{n}_{v,k+1}-\left[b(\hat{Q}^{p_1-n}_{h,k}-\hat{Q}^{p_1-n}_{v,k})-a(\hat{Q}^{p_1-n}_{h,k+1}+\hat{Q}^{p_1-n}_{v,k+1})+d(\hat{Q}^{p_2-n}_{h,k}-\hat{Q}^{p_2-n}_{v,k})-c(\hat{Q}^{p_2-n}_{h,k+1}+\hat{Q}^{p_2-n}_{v,k+1})\right]\,,
\end{aligned}
\label{dual-rail-transformed}
\end{equation}
\end{widetext}
where the temporal index $k$ was added as a label in all the operators. Lastly, as any linear combination of $X_{h,k}$ and $X_{v,k^\prime}$ is squeezed, we can define the following squeezed supermodes
\begin{widetext}
\begin{equation}
\begin{aligned}
X^+_k&=(X_{v,k-1}+X_{h,k})/2\\
&=\hat{Q}^{n}_{h,k}-\left[-a\hat{Q}^{p_1-n}_{v,k}-c\hat{Q}^{p_2-n}_{v,k}+ \frac{b}{2}(\hat{Q}^{p_1-n}_{h,k+1}+\hat{Q}^{p_1-n}_{v,k+1}+\hat{Q}^{p_1-n}_{h,k-1}-\hat{Q}^{p_1-n}_{v,k-1}) \right.+\\
&\quad\left.+\frac{d}{2}(\hat{Q}^{p_2-n}_{h,k+1}+\hat{Q}^{p_2-n}_{v,k+1}+\hat{Q}^{p_2-n}_{h,k-1}-\hat{Q}^{p_2-n}_{v,k-1})\right]\,,
\end{aligned}
\label{nullifier_+}
\end{equation}
\begin{equation}
\begin{aligned}
X^-_k&=(X_{v,k-1}-X_{h,k})/2\\
&=\hat{Q}^{n}_{v,k}-\left[-a\hat{Q}^{p_1-n}_{h,k} -c\hat{Q}^{p_2-n}_{h,k}+ \frac{b}{2}(-\hat{Q}^{p_1-n}_{h,k+1}-\hat{Q}^{p_1-n}_{v,k+1}+\hat{Q}^{p_1-n}_{h,k-1}-\hat{Q}^{p_1-n}_{v,k-1})  \right.+\\
&\quad\left.+\frac{d}{2}(-\hat{Q}^{p_2-n}_{h,k+1}-\hat{Q}^{p_2-n}_{v,k+1}+\hat{Q}^{p_2-n}_{h,k-1}-\hat{Q}^{p_2-n}_{v,k-1})\right]\,.
\end{aligned}
\label{nullifier_-}
\end{equation}
\end{widetext}
These quantities, after appropriate phase shifts, become the approximate nullifiers of the 2D cluster shown in Fig.\ref{Fig-2D-cluster}. 

\subsection{Time-varying pump structure}

The clusters generated by a time-varying pump structure can be obtained in a similar manner. For that purpose, we first add the temporal index $k$ to the cluster edges, i.e., 
\begin{equation}
\begin{aligned}
 a &\rightarrow a_k, \qquad b\rightarrow b_k,\\
 c &\rightarrow c_k, \qquad d\rightarrow d_k\,,
 \end{aligned}
 \end{equation} 
taking into account that the 1D clusters in different time bins can be different. We obtain the 2D cluster nullifiers by repeating the procedure leading to Eqs. \eqref{nullifier_+} and \eqref{nullifier_-} for a given pair of pump structures at the times $k$ and $k-1$. The result is in general a non-uniform 2D pattern, as presented in Fig.\ref{Fig-onthefly}.

\textit{Acknowledgments.}--This work was supported by Coordenação de Aperfeiçoamento de Pessoal de Nível Superior (CAPES), Fundação Carlos Chagas Filho de Amparo à Pesquisa do Estado do Rio de Janeiro (FAPERJ), Instituto Nacional de Ciência e Tecnologia de Informação Quântica (INCT-IQ) and Conselho Nacional de Desenvolvimento Científico e Tecnológico (CNPq). O.P. was supported by NSF grant PHY-1820882.

\bibliographystyle{apsrev4-2} 
\bibliography{references} 

\begin{thebibliography}{47}%
\makeatletter
\providecommand \@ifxundefined [1]{%
 \@ifx{#1\undefined}
}%
\providecommand \@ifnum [1]{%
 \ifnum #1\expandafter \@firstoftwo
 \else \expandafter \@secondoftwo
 \fi
}%
\providecommand \@ifx [1]{%
 \ifx #1\expandafter \@firstoftwo
 \else \expandafter \@secondoftwo
 \fi
}%
\providecommand \natexlab [1]{#1}%
\providecommand \enquote  [1]{``#1''}%
\providecommand \bibnamefont  [1]{#1}%
\providecommand \bibfnamefont [1]{#1}%
\providecommand \citenamefont [1]{#1}%
\providecommand \href@noop [0]{\@secondoftwo}%
\providecommand \href [0]{\begingroup \@sanitize@url \@href}%
\providecommand \@href[1]{\@@startlink{#1}\@@href}%
\providecommand \@@href[1]{\endgroup#1\@@endlink}%
\providecommand \@sanitize@url [0]{\catcode `\\12\catcode `\$12\catcode
  `\&12\catcode `\#12\catcode `\^12\catcode `\_12\catcode `\%12\relax}%
\providecommand \@@startlink[1]{}%
\providecommand \@@endlink[0]{}%
\providecommand \url  [0]{\begingroup\@sanitize@url \@url }%
\providecommand \@url [1]{\endgroup\@href {#1}{\urlprefix }}%
\providecommand \urlprefix  [0]{URL }%
\providecommand \Eprint [0]{\href }%
\providecommand \doibase [0]{https://doi.org/}%
\providecommand \selectlanguage [0]{\@gobble}%
\providecommand \bibinfo  [0]{\@secondoftwo}%
\providecommand \bibfield  [0]{\@secondoftwo}%
\providecommand \translation [1]{[#1]}%
\providecommand \BibitemOpen [0]{}%
\providecommand \bibitemStop [0]{}%
\providecommand \bibitemNoStop [0]{.\EOS\space}%
\providecommand \EOS [0]{\spacefactor3000\relax}%
\providecommand \BibitemShut  [1]{\csname bibitem#1\endcsname}%
\let\auto@bib@innerbib\@empty
\bibitem [{\citenamefont {Yokoyama}\ \emph {et~al.}(2013)\citenamefont
  {Yokoyama}, \citenamefont {Ukai}, \citenamefont {Armstrong}, \citenamefont
  {Sornphiphatphong}, \citenamefont {Kaji}, \citenamefont {Suzuki},
  \citenamefont {Yoshikawa}, \citenamefont {Yonezawa}, \citenamefont
  {Menicucci},\ and\ \citenamefont {Furusawa}}]{Yokoyama:13}%
  \BibitemOpen
  \bibfield  {author} {\bibinfo {author} {\bibfnamefont {S.}~\bibnamefont
  {Yokoyama}}, \bibinfo {author} {\bibfnamefont {R.}~\bibnamefont {Ukai}},
  \bibinfo {author} {\bibfnamefont {S.~C.}\ \bibnamefont {Armstrong}}, \bibinfo
  {author} {\bibfnamefont {C.}~\bibnamefont {Sornphiphatphong}}, \bibinfo
  {author} {\bibfnamefont {T.}~\bibnamefont {Kaji}}, \bibinfo {author}
  {\bibfnamefont {S.}~\bibnamefont {Suzuki}}, \bibinfo {author} {\bibfnamefont
  {J.-i.}\ \bibnamefont {Yoshikawa}}, \bibinfo {author} {\bibfnamefont
  {H.}~\bibnamefont {Yonezawa}}, \bibinfo {author} {\bibfnamefont {N.~C.}\
  \bibnamefont {Menicucci}},\ and\ \bibinfo {author} {\bibfnamefont
  {A.}~\bibnamefont {Furusawa}},\ }\href
  {https://doi.org/10.1038/nphoton.2013.287} {\bibfield  {journal} {\bibinfo
  {journal} {Nature Photonics}\ }\textbf {\bibinfo {volume} {7}},\ \bibinfo
  {pages} {982} (\bibinfo {year} {2013})}\BibitemShut {NoStop}%
\bibitem [{\citenamefont {Yoshikawa}\ \emph {et~al.}(2016)\citenamefont
  {Yoshikawa}, \citenamefont {Yokoyama}, \citenamefont {Kaji}, \citenamefont
  {Sornphiphatphong}, \citenamefont {Shiozawa}, \citenamefont {Makino},\ and\
  \citenamefont {Furusawa}}]{Yoshikawa:16}%
  \BibitemOpen
  \bibfield  {author} {\bibinfo {author} {\bibfnamefont {J.-i.}\ \bibnamefont
  {Yoshikawa}}, \bibinfo {author} {\bibfnamefont {S.}~\bibnamefont {Yokoyama}},
  \bibinfo {author} {\bibfnamefont {T.}~\bibnamefont {Kaji}}, \bibinfo {author}
  {\bibfnamefont {C.}~\bibnamefont {Sornphiphatphong}}, \bibinfo {author}
  {\bibfnamefont {Y.}~\bibnamefont {Shiozawa}}, \bibinfo {author}
  {\bibfnamefont {K.}~\bibnamefont {Makino}},\ and\ \bibinfo {author}
  {\bibfnamefont {A.}~\bibnamefont {Furusawa}},\ }\href
  {https://doi.org/10.1063/1.4962732} {\bibfield  {journal} {\bibinfo
  {journal} {APL Photonics}\ }\textbf {\bibinfo {volume} {1}},\ \bibinfo
  {pages} {060801} (\bibinfo {year} {2016})},\ \Eprint
  {https://arxiv.org/abs/https://doi.org/10.1063/1.4962732}
  {https://doi.org/10.1063/1.4962732} \BibitemShut {NoStop}%
\bibitem [{\citenamefont {Asavanant}\ \emph {et~al.}(2019)\citenamefont
  {Asavanant}, \citenamefont {Shiozawa}, \citenamefont {Yokoyama},
  \citenamefont {Charoensombutamon}, \citenamefont {Emura}, \citenamefont
  {Alexander}, \citenamefont {Takeda}, \citenamefont {Yoshikawa}, \citenamefont
  {Menicucci}, \citenamefont {Yonezawa},\ and\ \citenamefont
  {Furusawa}}]{Asavanant:19}%
  \BibitemOpen
  \bibfield  {author} {\bibinfo {author} {\bibfnamefont {W.}~\bibnamefont
  {Asavanant}}, \bibinfo {author} {\bibfnamefont {Y.}~\bibnamefont {Shiozawa}},
  \bibinfo {author} {\bibfnamefont {S.}~\bibnamefont {Yokoyama}}, \bibinfo
  {author} {\bibfnamefont {B.}~\bibnamefont {Charoensombutamon}}, \bibinfo
  {author} {\bibfnamefont {H.}~\bibnamefont {Emura}}, \bibinfo {author}
  {\bibfnamefont {R.~N.}\ \bibnamefont {Alexander}}, \bibinfo {author}
  {\bibfnamefont {S.}~\bibnamefont {Takeda}}, \bibinfo {author} {\bibfnamefont
  {J.-i.}\ \bibnamefont {Yoshikawa}}, \bibinfo {author} {\bibfnamefont {N.~C.}\
  \bibnamefont {Menicucci}}, \bibinfo {author} {\bibfnamefont {H.}~\bibnamefont
  {Yonezawa}},\ and\ \bibinfo {author} {\bibfnamefont {A.}~\bibnamefont
  {Furusawa}},\ }\href {https://doi.org/10.1126/science.aay2645} {\bibfield
  {journal} {\bibinfo  {journal} {Science}\ }\textbf {\bibinfo {volume}
  {366}},\ \bibinfo {pages} {373} (\bibinfo {year} {2019})}\BibitemShut
  {NoStop}%
\bibitem [{\citenamefont {Larsen}\ \emph {et~al.}(2019)\citenamefont {Larsen},
  \citenamefont {Guo}, \citenamefont {Breum}, \citenamefont
  {Neergaard-Nielsen},\ and\ \citenamefont {Andersen}}]{Larsen:19}%
  \BibitemOpen
  \bibfield  {author} {\bibinfo {author} {\bibfnamefont {M.~V.}\ \bibnamefont
  {Larsen}}, \bibinfo {author} {\bibfnamefont {X.}~\bibnamefont {Guo}},
  \bibinfo {author} {\bibfnamefont {C.~R.}\ \bibnamefont {Breum}}, \bibinfo
  {author} {\bibfnamefont {J.~S.}\ \bibnamefont {Neergaard-Nielsen}},\ and\
  \bibinfo {author} {\bibfnamefont {U.~L.}\ \bibnamefont {Andersen}},\ }\href
  {https://doi.org/10.1126/science.aay4354} {\bibfield  {journal} {\bibinfo
  {journal} {Science}\ }\textbf {\bibinfo {volume} {366}},\ \bibinfo {pages}
  {369} (\bibinfo {year} {2019})},\ \Eprint
  {https://arxiv.org/abs/https://science.sciencemag.org/content/366/6463/369.full.pdf}
  {https://science.sciencemag.org/content/366/6463/369.full.pdf} \BibitemShut
  {NoStop}%
\bibitem [{\citenamefont {Pysher}\ \emph {et~al.}(2011)\citenamefont {Pysher},
  \citenamefont {Miwa}, \citenamefont {Shahrokhshahi}, \citenamefont
  {Bloomer},\ and\ \citenamefont {Pfister}}]{Pysher:11}%
  \BibitemOpen
  \bibfield  {author} {\bibinfo {author} {\bibfnamefont {M.}~\bibnamefont
  {Pysher}}, \bibinfo {author} {\bibfnamefont {Y.}~\bibnamefont {Miwa}},
  \bibinfo {author} {\bibfnamefont {R.}~\bibnamefont {Shahrokhshahi}}, \bibinfo
  {author} {\bibfnamefont {R.}~\bibnamefont {Bloomer}},\ and\ \bibinfo {author}
  {\bibfnamefont {O.}~\bibnamefont {Pfister}},\ }\href
  {https://doi.org/10.1103/PhysRevLett.107.030505} {\bibfield  {journal}
  {\bibinfo  {journal} {Phys. Rev. Lett.}\ }\textbf {\bibinfo {volume} {107}},\
  \bibinfo {pages} {030505} (\bibinfo {year} {2011})}\BibitemShut {NoStop}%
\bibitem [{\citenamefont {Chen}\ \emph {et~al.}(2014)\citenamefont {Chen},
  \citenamefont {Menicucci},\ and\ \citenamefont {Pfister}}]{Chen:14}%
  \BibitemOpen
  \bibfield  {author} {\bibinfo {author} {\bibfnamefont {M.}~\bibnamefont
  {Chen}}, \bibinfo {author} {\bibfnamefont {N.~C.}\ \bibnamefont
  {Menicucci}},\ and\ \bibinfo {author} {\bibfnamefont {O.}~\bibnamefont
  {Pfister}},\ }\href {https://doi.org/10.1103/PhysRevLett.112.120505}
  {\bibfield  {journal} {\bibinfo  {journal} {Phys. Rev. Lett.}\ }\textbf
  {\bibinfo {volume} {112}},\ \bibinfo {pages} {120505} (\bibinfo {year}
  {2014})}\BibitemShut {NoStop}%
\bibitem [{\citenamefont {Menicucci}\ \emph {et~al.}(2008)\citenamefont
  {Menicucci}, \citenamefont {Flammia},\ and\ \citenamefont
  {Pfister}}]{Menicucci:08}%
  \BibitemOpen
  \bibfield  {author} {\bibinfo {author} {\bibfnamefont {N.~C.}\ \bibnamefont
  {Menicucci}}, \bibinfo {author} {\bibfnamefont {S.~T.}\ \bibnamefont
  {Flammia}},\ and\ \bibinfo {author} {\bibfnamefont {O.}~\bibnamefont
  {Pfister}},\ }\href {https://doi.org/10.1103/PhysRevLett.101.130501}
  {\bibfield  {journal} {\bibinfo  {journal} {Phys. Rev. Lett.}\ }\textbf
  {\bibinfo {volume} {101}},\ \bibinfo {pages} {130501} (\bibinfo {year}
  {2008})}\BibitemShut {NoStop}%
\bibitem [{\citenamefont {Zhu}\ \emph {et~al.}(2020)\citenamefont {Zhu},
  \citenamefont {Chang}, \citenamefont {González-Arciniegas}, \citenamefont
  {Pe'er}, \citenamefont {Higgins},\ and\ \citenamefont {Pfister}}]{Zhu:20}%
  \BibitemOpen
  \bibfield  {author} {\bibinfo {author} {\bibfnamefont {X.}~\bibnamefont
  {Zhu}}, \bibinfo {author} {\bibfnamefont {C.-H.}\ \bibnamefont {Chang}},
  \bibinfo {author} {\bibfnamefont {C.}~\bibnamefont {González-Arciniegas}},
  \bibinfo {author} {\bibfnamefont {A.}~\bibnamefont {Pe'er}}, \bibinfo
  {author} {\bibfnamefont {J.}~\bibnamefont {Higgins}},\ and\ \bibinfo {author}
  {\bibfnamefont {O.}~\bibnamefont {Pfister}},\ }\href@noop {} {\bibinfo
  {title} {Hypercubic cluster states in the phase modulated quantum optical
  frequency comb}} (\bibinfo {year} {2020}),\ \Eprint
  {https://arxiv.org/abs/1912.11215} {arXiv:1912.11215 [quant-ph]} \BibitemShut
  {NoStop}%
\bibitem [{\citenamefont {Roslund}\ \emph {et~al.}(2014)\citenamefont
  {Roslund}, \citenamefont {de~Ara{\'u}jo}, \citenamefont {Jiang},
  \citenamefont {Fabre},\ and\ \citenamefont {Treps}}]{Roslund2014}%
  \BibitemOpen
  \bibfield  {author} {\bibinfo {author} {\bibfnamefont {J.}~\bibnamefont
  {Roslund}}, \bibinfo {author} {\bibfnamefont {R.~M.}\ \bibnamefont
  {de~Ara{\'u}jo}}, \bibinfo {author} {\bibfnamefont {S.}~\bibnamefont
  {Jiang}}, \bibinfo {author} {\bibfnamefont {C.}~\bibnamefont {Fabre}},\ and\
  \bibinfo {author} {\bibfnamefont {N.}~\bibnamefont {Treps}},\ }\href
  {https://doi.org/10.1038/nphoton.2013.340} {\bibfield  {journal} {\bibinfo
  {journal} {Nature Photonics}\ }\textbf {\bibinfo {volume} {8}},\ \bibinfo
  {pages} {109} (\bibinfo {year} {2014})}\BibitemShut {NoStop}%
\bibitem [{\citenamefont {Cai}\ \emph {et~al.}(2017)\citenamefont {Cai},
  \citenamefont {Roslund}, \citenamefont {Ferrini}, \citenamefont {Arzani},
  \citenamefont {Xu}, \citenamefont {Fabre},\ and\ \citenamefont
  {Treps}}]{Cai2017}%
  \BibitemOpen
  \bibfield  {author} {\bibinfo {author} {\bibfnamefont {Y.}~\bibnamefont
  {Cai}}, \bibinfo {author} {\bibfnamefont {J.}~\bibnamefont {Roslund}},
  \bibinfo {author} {\bibfnamefont {G.}~\bibnamefont {Ferrini}}, \bibinfo
  {author} {\bibfnamefont {F.}~\bibnamefont {Arzani}}, \bibinfo {author}
  {\bibfnamefont {X.}~\bibnamefont {Xu}}, \bibinfo {author} {\bibfnamefont
  {C.}~\bibnamefont {Fabre}},\ and\ \bibinfo {author} {\bibfnamefont
  {N.}~\bibnamefont {Treps}},\ }\href {https://doi.org/10.1038/ncomms15645}
  {\bibfield  {journal} {\bibinfo  {journal} {Nature Communications}\ }\textbf
  {\bibinfo {volume} {8}},\ \bibinfo {pages} {15645} (\bibinfo {year}
  {2017})}\BibitemShut {NoStop}%
\bibitem [{\citenamefont {dos Santos}\ \emph {et~al.}(2009)\citenamefont {dos
  Santos}, \citenamefont {Dechoum},\ and\ \citenamefont
  {Khoury}}]{Coutinho:09}%
  \BibitemOpen
  \bibfield  {author} {\bibinfo {author} {\bibfnamefont {B.~C.}\ \bibnamefont
  {dos Santos}}, \bibinfo {author} {\bibfnamefont {K.}~\bibnamefont
  {Dechoum}},\ and\ \bibinfo {author} {\bibfnamefont {A.~Z.}\ \bibnamefont
  {Khoury}},\ }\href {https://doi.org/10.1103/PhysRevLett.103.230503}
  {\bibfield  {journal} {\bibinfo  {journal} {Phys. Rev. Lett.}\ }\textbf
  {\bibinfo {volume} {103}},\ \bibinfo {pages} {230503} (\bibinfo {year}
  {2009})}\BibitemShut {NoStop}%
\bibitem [{\citenamefont {Liu}\ \emph {et~al.}(2014)\citenamefont {Liu},
  \citenamefont {Guo}, \citenamefont {Cai}, \citenamefont {Guo},\ and\
  \citenamefont {Gao}}]{Liu:14}%
  \BibitemOpen
  \bibfield  {author} {\bibinfo {author} {\bibfnamefont {K.}~\bibnamefont
  {Liu}}, \bibinfo {author} {\bibfnamefont {J.}~\bibnamefont {Guo}}, \bibinfo
  {author} {\bibfnamefont {C.}~\bibnamefont {Cai}}, \bibinfo {author}
  {\bibfnamefont {S.}~\bibnamefont {Guo}},\ and\ \bibinfo {author}
  {\bibfnamefont {J.}~\bibnamefont {Gao}},\ }\href
  {https://doi.org/10.1103/PhysRevLett.113.170501} {\bibfield  {journal}
  {\bibinfo  {journal} {Phys. Rev. Lett.}\ }\textbf {\bibinfo {volume} {113}},\
  \bibinfo {pages} {170501} (\bibinfo {year} {2014})}\BibitemShut {NoStop}%
\bibitem [{\citenamefont {Liu}\ \emph {et~al.}(2016)\citenamefont {Liu},
  \citenamefont {Guo}, \citenamefont {Cai}, \citenamefont {Zhang},\ and\
  \citenamefont {Gao}}]{Liu:16}%
  \BibitemOpen
  \bibfield  {author} {\bibinfo {author} {\bibfnamefont {K.}~\bibnamefont
  {Liu}}, \bibinfo {author} {\bibfnamefont {J.}~\bibnamefont {Guo}}, \bibinfo
  {author} {\bibfnamefont {C.}~\bibnamefont {Cai}}, \bibinfo {author}
  {\bibfnamefont {J.}~\bibnamefont {Zhang}},\ and\ \bibinfo {author}
  {\bibfnamefont {J.}~\bibnamefont {Gao}},\ }\href
  {https://doi.org/10.1364/OL.41.005178} {\bibfield  {journal} {\bibinfo
  {journal} {Opt. Lett.}\ }\textbf {\bibinfo {volume} {41}},\ \bibinfo {pages}
  {5178} (\bibinfo {year} {2016})}\BibitemShut {NoStop}%
\bibitem [{\citenamefont {Cai}\ \emph {et~al.}(2018)\citenamefont {Cai},
  \citenamefont {Ma}, \citenamefont {Li}, \citenamefont {Guo}, \citenamefont
  {Liu}, \citenamefont {Sun}, \citenamefont {Yang},\ and\ \citenamefont
  {Gao}}]{Cai:18}%
  \BibitemOpen
  \bibfield  {author} {\bibinfo {author} {\bibfnamefont {C.}~\bibnamefont
  {Cai}}, \bibinfo {author} {\bibfnamefont {L.}~\bibnamefont {Ma}}, \bibinfo
  {author} {\bibfnamefont {J.}~\bibnamefont {Li}}, \bibinfo {author}
  {\bibfnamefont {H.}~\bibnamefont {Guo}}, \bibinfo {author} {\bibfnamefont
  {K.}~\bibnamefont {Liu}}, \bibinfo {author} {\bibfnamefont {H.}~\bibnamefont
  {Sun}}, \bibinfo {author} {\bibfnamefont {R.}~\bibnamefont {Yang}},\ and\
  \bibinfo {author} {\bibfnamefont {J.}~\bibnamefont {Gao}},\ }\href
  {https://doi.org/10.1364/PRJ.6.000479} {\bibfield  {journal} {\bibinfo
  {journal} {Photon. Res.}\ }\textbf {\bibinfo {volume} {6}},\ \bibinfo {pages}
  {479} (\bibinfo {year} {2018})}\BibitemShut {NoStop}%
\bibitem [{\citenamefont {Pooser}\ and\ \citenamefont
  {Jing}(2014)}]{Pooser:14}%
  \BibitemOpen
  \bibfield  {author} {\bibinfo {author} {\bibfnamefont {R.}~\bibnamefont
  {Pooser}}\ and\ \bibinfo {author} {\bibfnamefont {J.}~\bibnamefont {Jing}},\
  }\href {https://doi.org/10.1103/PhysRevA.90.043841} {\bibfield  {journal}
  {\bibinfo  {journal} {Phys. Rev. A}\ }\textbf {\bibinfo {volume} {90}},\
  \bibinfo {pages} {043841} (\bibinfo {year} {2014})}\BibitemShut {NoStop}%
\bibitem [{\citenamefont {Pan}\ \emph {et~al.}(2019)\citenamefont {Pan},
  \citenamefont {Yu}, \citenamefont {Zhou}, \citenamefont {Zhang},
  \citenamefont {Zhang}, \citenamefont {Lv}, \citenamefont {Li}, \citenamefont
  {Wang},\ and\ \citenamefont {Jing}}]{Pan:19}%
  \BibitemOpen
  \bibfield  {author} {\bibinfo {author} {\bibfnamefont {X.}~\bibnamefont
  {Pan}}, \bibinfo {author} {\bibfnamefont {S.}~\bibnamefont {Yu}}, \bibinfo
  {author} {\bibfnamefont {Y.}~\bibnamefont {Zhou}}, \bibinfo {author}
  {\bibfnamefont {K.}~\bibnamefont {Zhang}}, \bibinfo {author} {\bibfnamefont
  {K.}~\bibnamefont {Zhang}}, \bibinfo {author} {\bibfnamefont
  {S.}~\bibnamefont {Lv}}, \bibinfo {author} {\bibfnamefont {S.}~\bibnamefont
  {Li}}, \bibinfo {author} {\bibfnamefont {W.}~\bibnamefont {Wang}},\ and\
  \bibinfo {author} {\bibfnamefont {J.}~\bibnamefont {Jing}},\ }\href
  {https://doi.org/10.1103/PhysRevLett.123.070506} {\bibfield  {journal}
  {\bibinfo  {journal} {Phys. Rev. Lett.}\ }\textbf {\bibinfo {volume} {123}},\
  \bibinfo {pages} {070506} (\bibinfo {year} {2019})}\BibitemShut {NoStop}%
\bibitem [{\citenamefont {Li}\ \emph {et~al.}(2020)\citenamefont {Li},
  \citenamefont {Pan}, \citenamefont {Ren}, \citenamefont {Liu}, \citenamefont
  {Yu},\ and\ \citenamefont {Jing}}]{Li:20}%
  \BibitemOpen
  \bibfield  {author} {\bibinfo {author} {\bibfnamefont {S.}~\bibnamefont
  {Li}}, \bibinfo {author} {\bibfnamefont {X.}~\bibnamefont {Pan}}, \bibinfo
  {author} {\bibfnamefont {Y.}~\bibnamefont {Ren}}, \bibinfo {author}
  {\bibfnamefont {H.}~\bibnamefont {Liu}}, \bibinfo {author} {\bibfnamefont
  {S.}~\bibnamefont {Yu}},\ and\ \bibinfo {author} {\bibfnamefont
  {J.}~\bibnamefont {Jing}},\ }\href
  {https://doi.org/10.1103/PhysRevLett.124.083605} {\bibfield  {journal}
  {\bibinfo  {journal} {Phys. Rev. Lett.}\ }\textbf {\bibinfo {volume} {124}},\
  \bibinfo {pages} {083605} (\bibinfo {year} {2020})}\BibitemShut {NoStop}%
\bibitem [{\citenamefont {Zhang}\ \emph {et~al.}(2020)\citenamefont {Zhang},
  \citenamefont {Wang}, \citenamefont {Liu}, \citenamefont {Pan}, \citenamefont
  {Du}, \citenamefont {Lou}, \citenamefont {Yu}, \citenamefont {Lv},
  \citenamefont {Treps}, \citenamefont {Fabre},\ and\ \citenamefont
  {Jing}}]{Zhang:20}%
  \BibitemOpen
  \bibfield  {author} {\bibinfo {author} {\bibfnamefont {K.}~\bibnamefont
  {Zhang}}, \bibinfo {author} {\bibfnamefont {W.}~\bibnamefont {Wang}},
  \bibinfo {author} {\bibfnamefont {S.}~\bibnamefont {Liu}}, \bibinfo {author}
  {\bibfnamefont {X.}~\bibnamefont {Pan}}, \bibinfo {author} {\bibfnamefont
  {J.}~\bibnamefont {Du}}, \bibinfo {author} {\bibfnamefont {Y.}~\bibnamefont
  {Lou}}, \bibinfo {author} {\bibfnamefont {S.}~\bibnamefont {Yu}}, \bibinfo
  {author} {\bibfnamefont {S.}~\bibnamefont {Lv}}, \bibinfo {author}
  {\bibfnamefont {N.}~\bibnamefont {Treps}}, \bibinfo {author} {\bibfnamefont
  {C.}~\bibnamefont {Fabre}},\ and\ \bibinfo {author} {\bibfnamefont
  {J.}~\bibnamefont {Jing}},\ }\href
  {https://doi.org/10.1103/PhysRevLett.124.090501} {\bibfield  {journal}
  {\bibinfo  {journal} {Phys. Rev. Lett.}\ }\textbf {\bibinfo {volume} {124}},\
  \bibinfo {pages} {090501} (\bibinfo {year} {2020})}\BibitemShut {NoStop}%
\bibitem [{\citenamefont {Liu}\ \emph {et~al.}(2020)\citenamefont {Liu},
  \citenamefont {Lou},\ and\ \citenamefont {Jing}}]{Liu:20}%
  \BibitemOpen
  \bibfield  {author} {\bibinfo {author} {\bibfnamefont {S.}~\bibnamefont
  {Liu}}, \bibinfo {author} {\bibfnamefont {Y.}~\bibnamefont {Lou}},\ and\
  \bibinfo {author} {\bibfnamefont {J.}~\bibnamefont {Jing}},\ }\href
  {https://doi.org/10.1038/s41467-020-17616-4} {\bibfield  {journal} {\bibinfo
  {journal} {Nature Communications}\ }\textbf {\bibinfo {volume} {11}},\
  \bibinfo {pages} {3875} (\bibinfo {year} {2020})}\BibitemShut {NoStop}%
\bibitem [{\citenamefont {Wang}\ \emph {et~al.}(2020)\citenamefont {Wang},
  \citenamefont {Zhang},\ and\ \citenamefont {Jing}}]{Wang:20}%
  \BibitemOpen
  \bibfield  {author} {\bibinfo {author} {\bibfnamefont {W.}~\bibnamefont
  {Wang}}, \bibinfo {author} {\bibfnamefont {K.}~\bibnamefont {Zhang}},\ and\
  \bibinfo {author} {\bibfnamefont {J.}~\bibnamefont {Jing}},\ }\href
  {https://doi.org/10.1103/PhysRevLett.125.140501} {\bibfield  {journal}
  {\bibinfo  {journal} {Phys. Rev. Lett.}\ }\textbf {\bibinfo {volume} {125}},\
  \bibinfo {pages} {140501} (\bibinfo {year} {2020})}\BibitemShut {NoStop}%
\bibitem [{\citenamefont {Zhang}\ \emph {et~al.}(2017)\citenamefont {Zhang},
  \citenamefont {Wang}, \citenamefont {Yang}, \citenamefont {Liu},\ and\
  \citenamefont {Gao}}]{Zhang:17}%
  \BibitemOpen
  \bibfield  {author} {\bibinfo {author} {\bibfnamefont {J.}~\bibnamefont
  {Zhang}}, \bibinfo {author} {\bibfnamefont {J.~J.}\ \bibnamefont {Wang}},
  \bibinfo {author} {\bibfnamefont {R.~G.}\ \bibnamefont {Yang}}, \bibinfo
  {author} {\bibfnamefont {K.}~\bibnamefont {Liu}},\ and\ \bibinfo {author}
  {\bibfnamefont {J.~R.}\ \bibnamefont {Gao}},\ }\href
  {https://doi.org/10.1364/OE.25.027172} {\bibfield  {journal} {\bibinfo
  {journal} {Opt. Express}\ }\textbf {\bibinfo {volume} {25}},\ \bibinfo
  {pages} {27172} (\bibinfo {year} {2017})}\BibitemShut {NoStop}%
\bibitem [{\citenamefont {Yang}\ \emph {et~al.}(2020)\citenamefont {Yang},
  \citenamefont {Zhang}, \citenamefont {Klich}, \citenamefont
  {Gonz\'alez-Arciniegas},\ and\ \citenamefont {Pfister}}]{Yang:20}%
  \BibitemOpen
  \bibfield  {author} {\bibinfo {author} {\bibfnamefont {R.}~\bibnamefont
  {Yang}}, \bibinfo {author} {\bibfnamefont {J.}~\bibnamefont {Zhang}},
  \bibinfo {author} {\bibfnamefont {I.}~\bibnamefont {Klich}}, \bibinfo
  {author} {\bibfnamefont {C.}~\bibnamefont {Gonz\'alez-Arciniegas}},\ and\
  \bibinfo {author} {\bibfnamefont {O.}~\bibnamefont {Pfister}},\ }\href
  {https://doi.org/10.1103/PhysRevA.101.043832} {\bibfield  {journal} {\bibinfo
   {journal} {Phys. Rev. A}\ }\textbf {\bibinfo {volume} {101}},\ \bibinfo
  {pages} {043832} (\bibinfo {year} {2020})}\BibitemShut {NoStop}%
\bibitem [{\citenamefont {Patera}\ \emph {et~al.}(2012)\citenamefont {Patera},
  \citenamefont {Navarrete-Benlloch}, \citenamefont {de~Valc{\'a}rcel},\ and\
  \citenamefont {Fabre}}]{Patera:12}%
  \BibitemOpen
  \bibfield  {author} {\bibinfo {author} {\bibfnamefont {G.}~\bibnamefont
  {Patera}}, \bibinfo {author} {\bibfnamefont {C.}~\bibnamefont
  {Navarrete-Benlloch}}, \bibinfo {author} {\bibfnamefont {G.~J.}\ \bibnamefont
  {de~Valc{\'a}rcel}},\ and\ \bibinfo {author} {\bibfnamefont {C.}~\bibnamefont
  {Fabre}},\ }\href {https://doi.org/10.1140/epjd/e2012-30036-2} {\bibfield
  {journal} {\bibinfo  {journal} {The European Physical Journal D}\ }\textbf
  {\bibinfo {volume} {66}},\ \bibinfo {pages} {241} (\bibinfo {year}
  {2012})}\BibitemShut {NoStop}%
\bibitem [{\citenamefont {Menicucci}\ \emph {et~al.}(2007)\citenamefont
  {Menicucci}, \citenamefont {Flammia}, \citenamefont {Zaidi},\ and\
  \citenamefont {Pfister}}]{Menicucci:07}%
  \BibitemOpen
  \bibfield  {author} {\bibinfo {author} {\bibfnamefont {N.~C.}\ \bibnamefont
  {Menicucci}}, \bibinfo {author} {\bibfnamefont {S.~T.}\ \bibnamefont
  {Flammia}}, \bibinfo {author} {\bibfnamefont {H.}~\bibnamefont {Zaidi}},\
  and\ \bibinfo {author} {\bibfnamefont {O.}~\bibnamefont {Pfister}},\ }\href
  {https://doi.org/10.1103/PhysRevA.76.010302} {\bibfield  {journal} {\bibinfo
  {journal} {Phys. Rev. A}\ }\textbf {\bibinfo {volume} {76}},\ \bibinfo
  {pages} {010302} (\bibinfo {year} {2007})}\BibitemShut {NoStop}%
\bibitem [{\citenamefont {Menicucci}\ \emph {et~al.}(2011)\citenamefont
  {Menicucci}, \citenamefont {Flammia},\ and\ \citenamefont {van
  Loock}}]{Menicucci:11}%
  \BibitemOpen
  \bibfield  {author} {\bibinfo {author} {\bibfnamefont {N.~C.}\ \bibnamefont
  {Menicucci}}, \bibinfo {author} {\bibfnamefont {S.~T.}\ \bibnamefont
  {Flammia}},\ and\ \bibinfo {author} {\bibfnamefont {P.}~\bibnamefont {van
  Loock}},\ }\href {https://doi.org/10.1103/PhysRevA.83.042335} {\bibfield
  {journal} {\bibinfo  {journal} {Phys. Rev. A}\ }\textbf {\bibinfo {volume}
  {83}},\ \bibinfo {pages} {042335} (\bibinfo {year} {2011})}\BibitemShut
  {NoStop}%
\bibitem [{\citenamefont {Alves}\ \emph {et~al.}(2018)\citenamefont {Alves},
  \citenamefont {Barros}, \citenamefont {Tasca}, \citenamefont {Souza},\ and\
  \citenamefont {Khoury}}]{Alves2018}%
  \BibitemOpen
  \bibfield  {author} {\bibinfo {author} {\bibfnamefont {G.~B.}\ \bibnamefont
  {Alves}}, \bibinfo {author} {\bibfnamefont {R.~F.}\ \bibnamefont {Barros}},
  \bibinfo {author} {\bibfnamefont {D.~S.}\ \bibnamefont {Tasca}}, \bibinfo
  {author} {\bibfnamefont {C.~E.~R.}\ \bibnamefont {Souza}},\ and\ \bibinfo
  {author} {\bibfnamefont {A.~Z.}\ \bibnamefont {Khoury}},\ }\href
  {https://doi.org/10.1103/PhysRevA.98.063825} {\bibfield  {journal} {\bibinfo
  {journal} {Phys. Rev. A}\ }\textbf {\bibinfo {volume} {98}},\ \bibinfo
  {pages} {063825} (\bibinfo {year} {2018})}\BibitemShut {NoStop}%
\bibitem [{\citenamefont {Schwob}\ \emph {et~al.}(1998)\citenamefont {Schwob},
  \citenamefont {Cohadon}, \citenamefont {Fabre}, \citenamefont {Marte},
  \citenamefont {Ritsch}, \citenamefont {Gatti},\ and\ \citenamefont
  {Lugiato}}]{Schwob1998}%
  \BibitemOpen
  \bibfield  {author} {\bibinfo {author} {\bibfnamefont {C.}~\bibnamefont
  {Schwob}}, \bibinfo {author} {\bibfnamefont {P.}~\bibnamefont {Cohadon}},
  \bibinfo {author} {\bibfnamefont {C.}~\bibnamefont {Fabre}}, \bibinfo
  {author} {\bibfnamefont {M.}~\bibnamefont {Marte}}, \bibinfo {author}
  {\bibfnamefont {H.}~\bibnamefont {Ritsch}}, \bibinfo {author} {\bibfnamefont
  {A.}~\bibnamefont {Gatti}},\ and\ \bibinfo {author} {\bibfnamefont
  {L.}~\bibnamefont {Lugiato}},\ }\href {https://doi.org/10.1007/s003400050455}
  {\bibfield  {journal} {\bibinfo  {journal} {Applied Physics B}\ }\textbf
  {\bibinfo {volume} {66}},\ \bibinfo {pages} {685} (\bibinfo {year}
  {1998})}\BibitemShut {NoStop}%
\bibitem [{\citenamefont {Yariv}(1991)}]{Yariv}%
  \BibitemOpen
  \bibfield  {author} {\bibinfo {author} {\bibfnamefont {A.}~\bibnamefont
  {Yariv}},\ }\href@noop {} {\emph {\bibinfo {title} {Optical electronics}}},\
  \bibinfo {edition} {4th}\ ed.,\ HRW series in electrical engineering\
  (\bibinfo  {publisher} {Saunders College Publishing},\ \bibinfo {address}
  {Philadelphia, Pa. ; Sydney},\ \bibinfo {year} {1991})\BibitemShut {NoStop}%
\bibitem [{\citenamefont {Eckardt}\ \emph {et~al.}(1991)\citenamefont
  {Eckardt}, \citenamefont {Nabors}, \citenamefont {Kozlovsky},\ and\
  \citenamefont {Byer}}]{Eckardt:91}%
  \BibitemOpen
  \bibfield  {author} {\bibinfo {author} {\bibfnamefont {R.~C.}\ \bibnamefont
  {Eckardt}}, \bibinfo {author} {\bibfnamefont {C.~D.}\ \bibnamefont {Nabors}},
  \bibinfo {author} {\bibfnamefont {W.~J.}\ \bibnamefont {Kozlovsky}},\ and\
  \bibinfo {author} {\bibfnamefont {R.~L.}\ \bibnamefont {Byer}},\ }\href
  {https://doi.org/10.1364/JOSAB.8.000646} {\bibfield  {journal} {\bibinfo
  {journal} {J. Opt. Soc. Am. B}\ }\textbf {\bibinfo {volume} {8}},\ \bibinfo
  {pages} {646} (\bibinfo {year} {1991})}\BibitemShut {NoStop}%
\bibitem [{\citenamefont {Barros}\ \emph {et~al.}(2021)\citenamefont {Barros},
  \citenamefont {Alves},\ and\ \citenamefont {Khoury}}]{Barros:20}%
  \BibitemOpen
  \bibfield  {author} {\bibinfo {author} {\bibfnamefont {R.~F.}\ \bibnamefont
  {Barros}}, \bibinfo {author} {\bibfnamefont {G.~B.}\ \bibnamefont {Alves}},\
  and\ \bibinfo {author} {\bibfnamefont {A.~Z.}\ \bibnamefont {Khoury}},\
  }\href {https://doi.org/10.1103/PhysRevA.103.023511} {\bibfield  {journal}
  {\bibinfo  {journal} {Phys. Rev. A}\ }\textbf {\bibinfo {volume} {103}},\
  \bibinfo {pages} {023511} (\bibinfo {year} {2021})}\BibitemShut {NoStop}%
\bibitem [{\citenamefont {Wang}\ \emph {et~al.}(2014)\citenamefont {Wang},
  \citenamefont {Fan},\ and\ \citenamefont {Pfister}}]{Wang:14}%
  \BibitemOpen
  \bibfield  {author} {\bibinfo {author} {\bibfnamefont {P.}~\bibnamefont
  {Wang}}, \bibinfo {author} {\bibfnamefont {W.}~\bibnamefont {Fan}},\ and\
  \bibinfo {author} {\bibfnamefont {O.}~\bibnamefont {Pfister}},\ }\href@noop
  {} {\bibinfo {title} {Engineering large-scale entanglement in the quantum
  optical frequency comb: influence of the quasiphasematching bandwidth and of
  dispersion}} (\bibinfo {year} {2014}),\ \Eprint
  {https://arxiv.org/abs/1403.6631} {arXiv:1403.6631 [physics.optics]}
  \BibitemShut {NoStop}%
\bibitem [{\citenamefont {Gu}\ \emph {et~al.}(2009)\citenamefont {Gu},
  \citenamefont {Weedbrook}, \citenamefont {Menicucci}, \citenamefont {Ralph},\
  and\ \citenamefont {van Loock}}]{Gu:09}%
  \BibitemOpen
  \bibfield  {author} {\bibinfo {author} {\bibfnamefont {M.}~\bibnamefont
  {Gu}}, \bibinfo {author} {\bibfnamefont {C.}~\bibnamefont {Weedbrook}},
  \bibinfo {author} {\bibfnamefont {N.~C.}\ \bibnamefont {Menicucci}}, \bibinfo
  {author} {\bibfnamefont {T.~C.}\ \bibnamefont {Ralph}},\ and\ \bibinfo
  {author} {\bibfnamefont {P.}~\bibnamefont {van Loock}},\ }\href
  {https://doi.org/10.1103/PhysRevA.79.062318} {\bibfield  {journal} {\bibinfo
  {journal} {Phys. Rev. A}\ }\textbf {\bibinfo {volume} {79}},\ \bibinfo
  {pages} {062318} (\bibinfo {year} {2009})}\BibitemShut {NoStop}%
\bibitem [{\citenamefont {Alexander}\ \emph {et~al.}(2016)\citenamefont
  {Alexander}, \citenamefont {Wang}, \citenamefont {Sridhar}, \citenamefont
  {Chen}, \citenamefont {Pfister},\ and\ \citenamefont
  {Menicucci}}]{Alexander:16}%
  \BibitemOpen
  \bibfield  {author} {\bibinfo {author} {\bibfnamefont {R.~N.}\ \bibnamefont
  {Alexander}}, \bibinfo {author} {\bibfnamefont {P.}~\bibnamefont {Wang}},
  \bibinfo {author} {\bibfnamefont {N.}~\bibnamefont {Sridhar}}, \bibinfo
  {author} {\bibfnamefont {M.}~\bibnamefont {Chen}}, \bibinfo {author}
  {\bibfnamefont {O.}~\bibnamefont {Pfister}},\ and\ \bibinfo {author}
  {\bibfnamefont {N.~C.}\ \bibnamefont {Menicucci}},\ }\href
  {https://doi.org/10.1103/PhysRevA.94.032327} {\bibfield  {journal} {\bibinfo
  {journal} {Phys. Rev. A}\ }\textbf {\bibinfo {volume} {94}},\ \bibinfo
  {pages} {032327} (\bibinfo {year} {2016})}\BibitemShut {NoStop}%
\bibitem [{\citenamefont {Menicucci}(2014)}]{Menicucci:14}%
  \BibitemOpen
  \bibfield  {author} {\bibinfo {author} {\bibfnamefont {N.~C.}\ \bibnamefont
  {Menicucci}},\ }\href {https://doi.org/10.1103/PhysRevLett.112.120504}
  {\bibfield  {journal} {\bibinfo  {journal} {Phys. Rev. Lett.}\ }\textbf
  {\bibinfo {volume} {112}},\ \bibinfo {pages} {120504} (\bibinfo {year}
  {2014})}\BibitemShut {NoStop}%
\bibitem [{\citenamefont {Lloyd}\ and\ \citenamefont
  {Braunstein}(1999)}]{Lloyd:99}%
  \BibitemOpen
  \bibfield  {author} {\bibinfo {author} {\bibfnamefont {S.}~\bibnamefont
  {Lloyd}}\ and\ \bibinfo {author} {\bibfnamefont {S.~L.}\ \bibnamefont
  {Braunstein}},\ }\href {https://doi.org/10.1103/PhysRevLett.82.1784}
  {\bibfield  {journal} {\bibinfo  {journal} {Phys. Rev. Lett.}\ }\textbf
  {\bibinfo {volume} {82}},\ \bibinfo {pages} {1784} (\bibinfo {year}
  {1999})}\BibitemShut {NoStop}%
\bibitem [{\citenamefont {Braunstein}\ and\ \citenamefont {van
  Loock}(2005)}]{Braunstein:05}%
  \BibitemOpen
  \bibfield  {author} {\bibinfo {author} {\bibfnamefont {S.~L.}\ \bibnamefont
  {Braunstein}}\ and\ \bibinfo {author} {\bibfnamefont {P.}~\bibnamefont {van
  Loock}},\ }\href {https://doi.org/10.1103/RevModPhys.77.513} {\bibfield
  {journal} {\bibinfo  {journal} {Rev. Mod. Phys.}\ }\textbf {\bibinfo {volume}
  {77}},\ \bibinfo {pages} {513} (\bibinfo {year} {2005})}\BibitemShut
  {NoStop}%
\bibitem [{\citenamefont {Sasada}\ and\ \citenamefont
  {Okamoto}(2003)}]{Sasada:03}%
  \BibitemOpen
  \bibfield  {author} {\bibinfo {author} {\bibfnamefont {H.}~\bibnamefont
  {Sasada}}\ and\ \bibinfo {author} {\bibfnamefont {M.}~\bibnamefont
  {Okamoto}},\ }\href {https://doi.org/10.1103/PhysRevA.68.012323} {\bibfield
  {journal} {\bibinfo  {journal} {Phys. Rev. A}\ }\textbf {\bibinfo {volume}
  {68}},\ \bibinfo {pages} {012323} (\bibinfo {year} {2003})}\BibitemShut
  {NoStop}%
\bibitem [{\citenamefont {Passos}\ \emph {et~al.}(2020)\citenamefont {Passos},
  \citenamefont {Junior}, \citenamefont {de~Oliveira}, \citenamefont {Khoury},\
  and\ \citenamefont {Huguenin}}]{Passos:20}%
  \BibitemOpen
  \bibfield  {author} {\bibinfo {author} {\bibfnamefont {M.~H.~M.}\
  \bibnamefont {Passos}}, \bibinfo {author} {\bibfnamefont {A.~d.~O.}\
  \bibnamefont {Junior}}, \bibinfo {author} {\bibfnamefont {M.~C.}\
  \bibnamefont {de~Oliveira}}, \bibinfo {author} {\bibfnamefont {A.~Z.}\
  \bibnamefont {Khoury}},\ and\ \bibinfo {author} {\bibfnamefont {J.~A.~O.}\
  \bibnamefont {Huguenin}},\ }\href
  {https://doi.org/10.1103/PhysRevA.102.062601} {\bibfield  {journal} {\bibinfo
   {journal} {Phys. Rev. A}\ }\textbf {\bibinfo {volume} {102}},\ \bibinfo
  {pages} {062601} (\bibinfo {year} {2020})}\BibitemShut {NoStop}%
\bibitem [{\citenamefont {Van~den Nest}\ \emph {et~al.}(2006)\citenamefont
  {Van~den Nest}, \citenamefont {Miyake}, \citenamefont {D\"ur},\ and\
  \citenamefont {Briegel}}]{Nest:06}%
  \BibitemOpen
  \bibfield  {author} {\bibinfo {author} {\bibfnamefont {M.}~\bibnamefont
  {Van~den Nest}}, \bibinfo {author} {\bibfnamefont {A.}~\bibnamefont
  {Miyake}}, \bibinfo {author} {\bibfnamefont {W.}~\bibnamefont {D\"ur}},\ and\
  \bibinfo {author} {\bibfnamefont {H.~J.}\ \bibnamefont {Briegel}},\ }\href
  {https://doi.org/10.1103/PhysRevLett.97.150504} {\bibfield  {journal}
  {\bibinfo  {journal} {Phys. Rev. Lett.}\ }\textbf {\bibinfo {volume} {97}},\
  \bibinfo {pages} {150504} (\bibinfo {year} {2006})}\BibitemShut {NoStop}%
\bibitem [{\citenamefont {Martinelli}\ \emph {et~al.}(2004)\citenamefont
  {Martinelli}, \citenamefont {Huguenin}, \citenamefont {Nussenzveig},\ and\
  \citenamefont {Khoury}}]{Martinelli:04}%
  \BibitemOpen
  \bibfield  {author} {\bibinfo {author} {\bibfnamefont {M.}~\bibnamefont
  {Martinelli}}, \bibinfo {author} {\bibfnamefont {J.~A.~O.}\ \bibnamefont
  {Huguenin}}, \bibinfo {author} {\bibfnamefont {P.}~\bibnamefont
  {Nussenzveig}},\ and\ \bibinfo {author} {\bibfnamefont {A.~Z.}\ \bibnamefont
  {Khoury}},\ }\href {https://doi.org/10.1103/PhysRevA.70.013812} {\bibfield
  {journal} {\bibinfo  {journal} {Phys. Rev. A}\ }\textbf {\bibinfo {volume}
  {70}},\ \bibinfo {pages} {013812} (\bibinfo {year} {2004})}\BibitemShut
  {NoStop}%
\bibitem [{\citenamefont {Shaked}\ \emph {et~al.}(2018)\citenamefont {Shaked},
  \citenamefont {Michael}, \citenamefont {Vered}, \citenamefont {Bello},
  \citenamefont {Rosenbluh},\ and\ \citenamefont {Pe'er}}]{Shaked:18}%
  \BibitemOpen
  \bibfield  {author} {\bibinfo {author} {\bibfnamefont {Y.}~\bibnamefont
  {Shaked}}, \bibinfo {author} {\bibfnamefont {Y.}~\bibnamefont {Michael}},
  \bibinfo {author} {\bibfnamefont {R.~Z.}\ \bibnamefont {Vered}}, \bibinfo
  {author} {\bibfnamefont {L.}~\bibnamefont {Bello}}, \bibinfo {author}
  {\bibfnamefont {M.}~\bibnamefont {Rosenbluh}},\ and\ \bibinfo {author}
  {\bibfnamefont {A.}~\bibnamefont {Pe'er}},\ }\href
  {https://doi.org/10.1038/s41467-018-03083-5} {\bibfield  {journal} {\bibinfo
  {journal} {Nature Communications}\ }\textbf {\bibinfo {volume} {9}},\
  \bibinfo {pages} {609} (\bibinfo {year} {2018})}\BibitemShut {NoStop}%
\bibitem [{\citenamefont {Morizur}\ \emph {et~al.}(2010)\citenamefont
  {Morizur}, \citenamefont {Nicholls}, \citenamefont {Jian}, \citenamefont
  {Armstrong}, \citenamefont {Treps}, \citenamefont {Hage}, \citenamefont
  {Hsu}, \citenamefont {Bowen}, \citenamefont {Janousek},\ and\ \citenamefont
  {Bachor}}]{Morizur:10}%
  \BibitemOpen
  \bibfield  {author} {\bibinfo {author} {\bibfnamefont {J.-F.}\ \bibnamefont
  {Morizur}}, \bibinfo {author} {\bibfnamefont {L.}~\bibnamefont {Nicholls}},
  \bibinfo {author} {\bibfnamefont {P.}~\bibnamefont {Jian}}, \bibinfo {author}
  {\bibfnamefont {S.}~\bibnamefont {Armstrong}}, \bibinfo {author}
  {\bibfnamefont {N.}~\bibnamefont {Treps}}, \bibinfo {author} {\bibfnamefont
  {B.}~\bibnamefont {Hage}}, \bibinfo {author} {\bibfnamefont {M.}~\bibnamefont
  {Hsu}}, \bibinfo {author} {\bibfnamefont {W.}~\bibnamefont {Bowen}}, \bibinfo
  {author} {\bibfnamefont {J.}~\bibnamefont {Janousek}},\ and\ \bibinfo
  {author} {\bibfnamefont {H.-A.}\ \bibnamefont {Bachor}},\ }\href
  {https://doi.org/10.1364/JOSAA.27.002524} {\bibfield  {journal} {\bibinfo
  {journal} {J. Opt. Soc. Am. A}\ }\textbf {\bibinfo {volume} {27}},\ \bibinfo
  {pages} {2524} (\bibinfo {year} {2010})}\BibitemShut {NoStop}%
\bibitem [{\citenamefont {Fontaine}\ \emph {et~al.}(2019)\citenamefont
  {Fontaine}, \citenamefont {Ryf}, \citenamefont {Chen}, \citenamefont
  {Neilson}, \citenamefont {Kim},\ and\ \citenamefont
  {Carpenter}}]{Fontaine:19}%
  \BibitemOpen
  \bibfield  {author} {\bibinfo {author} {\bibfnamefont {N.~K.}\ \bibnamefont
  {Fontaine}}, \bibinfo {author} {\bibfnamefont {R.}~\bibnamefont {Ryf}},
  \bibinfo {author} {\bibfnamefont {H.}~\bibnamefont {Chen}}, \bibinfo {author}
  {\bibfnamefont {D.~T.}\ \bibnamefont {Neilson}}, \bibinfo {author}
  {\bibfnamefont {K.}~\bibnamefont {Kim}},\ and\ \bibinfo {author}
  {\bibfnamefont {J.}~\bibnamefont {Carpenter}},\ }\href
  {https://doi.org/10.1038/s41467-019-09840-4} {\bibfield  {journal} {\bibinfo
  {journal} {Nature Communications}\ }\textbf {\bibinfo {volume} {10}},\
  \bibinfo {pages} {1865} (\bibinfo {year} {2019})}\BibitemShut {NoStop}%
\bibitem [{\citenamefont {Fickler}\ \emph {et~al.}(2020)\citenamefont
  {Fickler}, \citenamefont {Bouchard}, \citenamefont {Giese}, \citenamefont
  {Grillo}, \citenamefont {Leuchs},\ and\ \citenamefont {Karimi}}]{Fickler:20}%
  \BibitemOpen
  \bibfield  {author} {\bibinfo {author} {\bibfnamefont {R.}~\bibnamefont
  {Fickler}}, \bibinfo {author} {\bibfnamefont {F.}~\bibnamefont {Bouchard}},
  \bibinfo {author} {\bibfnamefont {E.}~\bibnamefont {Giese}}, \bibinfo
  {author} {\bibfnamefont {V.}~\bibnamefont {Grillo}}, \bibinfo {author}
  {\bibfnamefont {G.}~\bibnamefont {Leuchs}},\ and\ \bibinfo {author}
  {\bibfnamefont {E.}~\bibnamefont {Karimi}},\ }\href
  {https://doi.org/10.1088/2040-8986/ab6303} {\bibfield  {journal} {\bibinfo
  {journal} {Journal of Optics}\ }\textbf {\bibinfo {volume} {22}},\ \bibinfo
  {pages} {024001} (\bibinfo {year} {2020})}\BibitemShut {NoStop}%
\bibitem [{\citenamefont {Simon}\ \emph {et~al.}(1988)\citenamefont {Simon},
  \citenamefont {Sudarshan},\ and\ \citenamefont {Mukunda}}]{Simon:88}%
  \BibitemOpen
  \bibfield  {author} {\bibinfo {author} {\bibfnamefont {R.}~\bibnamefont
  {Simon}}, \bibinfo {author} {\bibfnamefont {E.~C.~G.}\ \bibnamefont
  {Sudarshan}},\ and\ \bibinfo {author} {\bibfnamefont {N.}~\bibnamefont
  {Mukunda}},\ }\href {https://doi.org/10.1103/PhysRevA.37.3028} {\bibfield
  {journal} {\bibinfo  {journal} {Phys. Rev. A}\ }\textbf {\bibinfo {volume}
  {37}},\ \bibinfo {pages} {3028} (\bibinfo {year} {1988})}\BibitemShut
  {NoStop}%
\bibitem [{\citenamefont {Simon}(2000)}]{Simon:00}%
  \BibitemOpen
  \bibfield  {author} {\bibinfo {author} {\bibfnamefont {R.}~\bibnamefont
  {Simon}},\ }\href {https://doi.org/10.1103/PhysRevLett.84.2726} {\bibfield
  {journal} {\bibinfo  {journal} {Phys. Rev. Lett.}\ }\textbf {\bibinfo
  {volume} {84}},\ \bibinfo {pages} {2726} (\bibinfo {year}
  {2000})}\BibitemShut {NoStop}%
\bibitem [{\citenamefont {Peres}(1996)}]{Peres:96}%
  \BibitemOpen
  \bibfield  {author} {\bibinfo {author} {\bibfnamefont {A.}~\bibnamefont
  {Peres}},\ }\href {https://doi.org/10.1103/PhysRevLett.77.1413} {\bibfield
  {journal} {\bibinfo  {journal} {Phys. Rev. Lett.}\ }\textbf {\bibinfo
  {volume} {77}},\ \bibinfo {pages} {1413} (\bibinfo {year}
  {1996})}\BibitemShut {NoStop}%
\end{thebibliography}%

\end{document}